\documentclass[sigconf]{acmart}

\usepackage{tabularx}
\usepackage{multirow}

\usepackage[most]{tcolorbox}
\usepackage{soul}

\newtcolorbox[auto counter]{transcriptbox}[2][]{
    enhanced,
    colback=gray!5,
    colframe=black!40,
    boxrule=0.5pt,
    arc=2pt,
    left=6pt,
    right=6pt,
    top=6pt,
    bottom=6pt,
    fonttitle=\bfseries,
    fontupper=\itshape,
    title={Transcript~\thetcbcounter: #2},
    #1
}

\AtBeginDocument{%
  }

\setcopyright{acmlicensed}
\copyrightyear{2018}
\acmYear{2018}
\acmDOI{XXXXXXX.XXXXXXX}
\acmConference[Conference acronym 'XX]{Make sure to enter the correct
  conference title from your rights confirmation email}{June 03--05,
  2018}{Woodstock, NY}
\acmISBN{978-1-4503-XXXX-X/2018/06}

\usepackage{tikz}
\usepackage{float}

\usepackage{xspace}
\newcommand{\parabold}[1]{%
  \par\vspace{1ex}
  \noindent\textbf{#1. }%
}
\newcommand{\subsubheading}[1]{\subsubsection{\textbf{#1}}}

\usepackage{listings}
\lstdefinestyle{promptstyle}{%
  basicstyle=\ttfamily\scriptsize,
  breaklines=true,
  breakautoindent=false,
  breakindent=0pt,
  columns=fullflexible,
  keepspaces=true,
  showstringspaces=false,
  frame=single,
  framesep=4pt,
  rulecolor=\color{black!30},
  xleftmargin=2pt,
  xrightmargin=2pt,
  aboveskip=6pt,
  belowskip=6pt,
}

\begin{document}
\sethlcolor{blue!15}
\newcommand{\verbinline}[1]{\hl{\texttt{\,#1\,}}}

\title[Written Response Vs Dialogic Activities with AI]{From Written Response to Dialogue with AI: How Activity Format, Interaction Modality, and Language Impact Student Learning and Engagement}

\author{Deepak Varuvel Dennison}
\orcid{0009-0004-8355-5024}
\affiliation{%
  \institution{Cornell University}
  \city{Ithaca}
  \state{New York}
  \country{USA}
}
\email{dv292@cornell.edu}

\author{Connie Zhang}
\affiliation{%
  \institution{Cornell University}
  \city{Ithaca}
  \state{New York}
  \country{USA}
}
\email{cz467@cornell.edu}

\author{Rene Kizilcec}
\affiliation{%
  \institution{Cornell University}
  \city{Ithaca}
  \state{New York}
  \country{USA}
}
\email{kizilcec@cornell.edu}

\author{Aditya Vashistha}
\affiliation{%
  \institution{Cornell University}
  \city{Ithaca}
  \state{New York}
  \country{USA}
}
\email{adityav@cornell.edu}

\renewcommand{\shortauthors}{Trovato et al.}

\begin{abstract}
The widespread availability of LLMs is challenging written learning activities, as students can increasingly generate responses without necessarily engaging with the learning content. Conversational AI creates an opportunity to redesign these activities as dialogue, while multilingual capabilities may make such dialogue more accessible to students learning through a non-native language. We conducted a field study with 305 native Kannada-speaking undergraduate students at English-medium institutions in Karnataka, India. We compared written response activities with text- and voice-based dialogic activities with AI, each conducted in English-only or bilingual Kannada–English settings. Students who completed dialogic activities spent more time on the activities, contributed more, and reported greater interest and self-efficacy than those completing written responses, although fewer students completed the dialogic activities overall. Knowledge increased across all conditions, with no reliable differences in gains between activity formats or languages. Language shaped participation differently across modalities: bilingual interaction was particularly beneficial in voice dialogue, where it reduced articulation difficulties, increased turns demonstrating understanding, and reduced conversation abandonment. However, students also valued English because of its connection to their academic and professional aspirations. These findings show that designing dialogic learning with AI requires more than choosing between writing and dialogue, voice and text, or English and students’ native languages. We highlight opportunities to give learners greater control over modality, information, and pace, and to use native languages as translanguaging support rather than as a replacement for English.
\end{abstract}

\ccsdesc[500]{Do Not Use This Code~Generate the Correct Terms for Your Paper}
\ccsdesc[300]{Do Not Use This Code~Generate the Correct Terms for Your Paper}
\ccsdesc{Do Not Use This Code~Generate the Correct Terms for Your Paper}
\ccsdesc[100]{Do Not Use This Code~Generate the Correct Terms for Your Paper}

\keywords{Do, Not, Use, This, Code, Put, the, Correct, Terms, for,
  Your, Paper}
\begin{teaserfigure}
  \includegraphics[width=\textwidth]{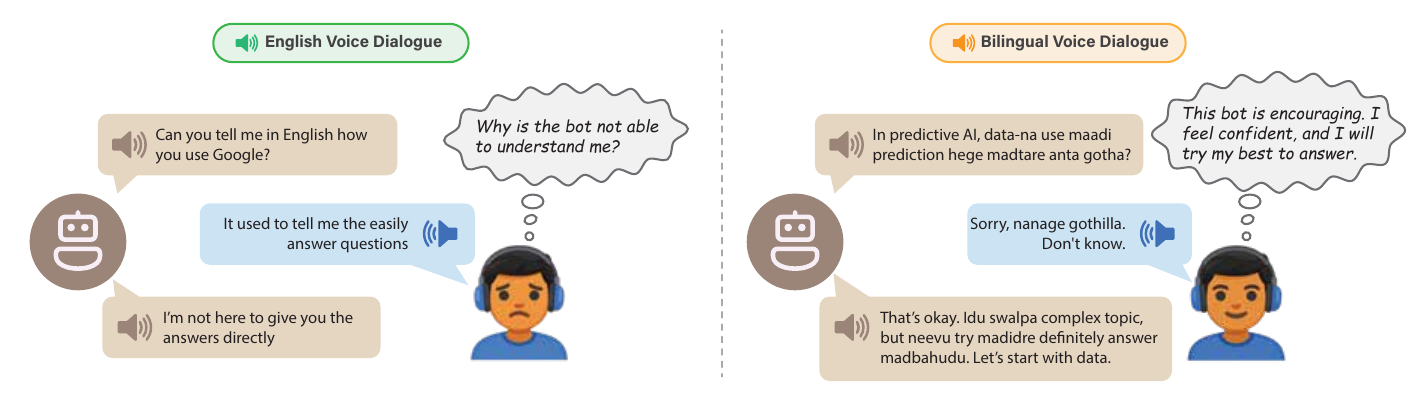}
  \caption{Examples of voice dialogue in the English-only (left) and bilingual Kannada--English (right) conditions.}
\Description{Two example conversations between students and an AI tutor. On the left, an English-only exchange shows a student attempting to explain how Google answers questions; the tutor repeatedly misinterprets the student's response as a request for a direct answer, resulting in a communication breakdown. On the right, a bilingual Kannada--English exchange shows a student saying they do not know the answer to a question about predictive AI. The tutor reassures the student, acknowledges that the topic is complex, and guides them to begin reasoning about the role of data.}
  \label{fig:teaser}
\end{teaserfigure}

\received{20 February 2007}
\received[revised]{12 March 2009}
\received[accepted]{5 June 2009}

\maketitle

\section{Introduction}

Written activities, where students are prompted to write a response based on their learning, have long been a staple in education \cite{ackermanPromiseWritingLearn1993, kleinReopeningInquiryCognitive1999, kimEmpoweringDiverseLearners2024}. However, the widespread availability of Large Language Models (LLMs) is challenging the effectiveness of such activities. Students can now use LLMs to produce written responses without any cognitive engagement with the learning content, making it difficult to determine whether completing a writing activity reflects meaningful learning \cite{marinaDetectionRethinkingEducation2025, wattsComparingStudentGenerative2023}. At the same time, conversational AI systems based on LLMs have created opportunities to redesign written activities as dialogic learning activities, in which students develop and articulate their thinking through an ongoing dialogue involving feedback, follow-up questions, and reflection \cite{bealeDialogicPedagogyLarge2025, tianSystematicLiteratureReview2025}. 

Conversational AI also creates opportunities to make dialogic learning more accessible to students learning in a non-native language. This is particularly important in English-medium instruction (EMI) settings, whose adoption has expanded rapidly across the world \cite{thakurEnglishMediumBoom2025, macaroSystematicReviewEnglish2018, fenton-smithUnpackingEnglishMedium2026}, driven by the educational and economic opportunities associated with English \cite{bhattacharyaMediatingInequalitiesExploring2013, khanEnglishLinguisticCapital2025, azamReturnsEnglishLanguageSkills}. Although EMI provides access to these opportunities, learning academic concepts through a non-native language can pose substantial challenges. Without adequate support, it increases cognitive demands and hinder comprehension, academic achievement, and conceptual understanding \cite{karabayEvolutionEnglishMedium2024, rousselLearningSubjectContent2017a, treffers-dallerHowReadyAre2022, civanEffectMediumInstruction2016, maInvestigationImpactCognitive2025}.
On the other hand, prior work shows that bilingual support can reduce linguistic barriers and support deeper conceptual understanding \cite{alhazmiImpactBilingualVs2025a, rousselAdvantagesListeningAcademic2022, rocabadoYouCanStand2025}. 
However, providing bilingual support remains difficult in practice. Teachers often lack the time, linguistic capacity, or resources required to support students with diverse language needs \cite{brisk2012challenges}.

Multilingual conversational AI offers a way to combine these opportunities by transforming written activities into scalable, bilingual dialogic learning activities. AI can engage students in dialogue while incorporating both English and their native languages. However, little is known about how replacing written activities with AI-based dialogic activities affects learning and engagement in combination with factors such as interaction modality (text or voice) and language use (English-only or bilingual).

To address this gap, we conducted a field study with 305 native Kannada-speaking undergraduate students at English-medium institutions in Karnataka, India. We compare written activities with AI-based dialogic activities across text and voice modalities and English-only and bilingual Kannada--English settings, examining how dialogue, modality, and language shape engagement and learning. Specifically, we ask:

\begin{itemize}

\item[\textbf{RQ1:}] How do dialogic activities with AI compare with written activities in students' engagement, completion, knowledge gains, interest, and self-efficacy?

\item[\textbf{RQ2:}] In dialogic activities with AI, how do interaction modality (text vs voice) and language (English-only vs bilingual) affect students' engagement and learning?

\item[\textbf{RQ3:}] What factors shape students' experiences and participation in dialogic activities with AI?

\end{itemize}

Our findings show that compared to written activities, AI-based dialogic activities meaningfully changed how students engaged with learning. AI-based dialogue produced learning gains comparable to those of written activities while also increasing students’ engagement, interest, and self-efficacy. Students valued conversational AI for providing immediate feedback, hint, and opportunities to ask follow-up questions without fear of judgment. Language and modality further shaped participation. Bilingual support was especially helpful in voice conversations, where it encouraged students to contribute more substantively. Although students found bilingual support easier to use for learning, many preferred interacting in English to improve their language skills, creating a tension between immediate accessibility and longer-term goals. Overall, these findings highlight the importance of designing dialogic activities with AI that supports both learners' current language abilities and their longer-term aspirations.

In summary, this paper makes three contributions:

\begin{itemize}

\item We show, through a field deployment with 305 undergraduates in English-medium institutions in Karnataka, India, that converting a written learning activity into AI-based dialogic activity substantially increases engagement, interest, and self-efficacy while producing comparable knowledge gains, though it lowers overall completion. We therefore demonstrate a tradeoff between engagement and completion --- although students who completed dialogic activities were more engaged, fewer students completed them overall.

\item We show that the effects of language depended on the interaction modality: Bilingual interaction improved the quality of participation in voice, whereas these language differences were not evident in text. Compared with English-only voice interaction, bilingual voice reduced articulation difficulties, increased turns demonstrating understanding, and reduced conversation abandonment. It also increased students' interest and self-efficacy. Overall, bilingual interaction changed the quality of students' participation rather than simply increasing how much they participated.

\item We identify a tension between making dialogic activities easier to participate in and supporting students' goals of improving their English. While bilingual interaction helps students express themselves more easily, many students also prefer to use English because they want to improve their English proficiency. We translate this tension into two design directions: giving learners more control over modality, information persistence, and turn-taking pace, and using students' native languages as translanguaging support for learning rather than as a replacement for English.

\end{itemize}

\section{Related Work}

Our work brings together three strands of research: the shift from written responses to AI-based dialogic learning, the role of interaction modality in shaping dialogue, and the use of bilingual support for learning academic content in English-medium instruction (EMI). 

\subsection{From Written Responses to Dialogue with AI}

Written activities have long been used to help students learn by requiring them to retrieve, organize, and articulate their understanding \citep{ackermanPromiseWritingLearn1993, kleinReopeningInquiryCognitive1999, kimEmpoweringDiverseLearners2024}. However, for the writing activities to be effective, students have to think themselves and write their own response. With the widespread availability of LLMs, students can now generate written answers without necessarily engaging in the thinking process \citep{marinaDetectionRethinkingEducation2025, wattsComparingStudentGenerative2023}. This development raises a broader question about whether written response activities remain an effective way to elicit and sustain students' thinking. Through their large-scale study, \citet{chirikovGenerativeAIUse2026} have shown how generative AI has made higher education assessments less reliable as evidence of students' capability and has called for a wide discipline-specific assessment reform. 

Dialogic learning offers one way to rethink these activities. Rather than asking students to produce a single finished written response, dialogic activities develop understanding through dialogue in which learners articulate ideas, receive feedback, and revise their thinking \cite{renshawDialogicLearningTeaching2004}. Interactive one-to-one tutoring, a form of dialogic learning, is considered one of the most effective forms of instruction \citep{chiLearningHumanTutoring2001}, and researchers have sought to reproduce its benefits through intelligent tutoring systems and conversational agents \citep{graesserCollaborativeDialoguePatterns1995, koedinger2001cognitive, paladinesSystematicLiteratureReview2020}. Conversational AI extends this line of work by making it possible to facilitate dialogic learning at scale \citep{bealeDialogicPedagogyLarge2025, tianSystematicLiteratureReview2025}.

Recent studies illustrate how conversational AI capabilities can support learning. Learners have valued adaptive follow-up questions that prompted deeper reflection, as well as continuous and personalized feedback that allowed them to disagree or express uncertainty without fear of judgment \citep{daryantoConversateSupportingReflective2025}. When positioned as a dialogic partner rather than an authoritative source, AI has also elicited perspective-taking, reasoning, argumentation, and creative thinking \citep{tangGenerativeAIDialogic2026}. More broadly, emerging evidence suggests that AI-based dialogic activities can support engagement and learning at scale \citep{tianSystematicLiteratureReview2025, liuAmIWriteExploringScalable2026}.

Although prior work shows the promise of AI-supported dialogue, little is known about how replacing a written response with dialogic activity with AI affects learning and engagement within the same academic context. We address this gap by comparing the two formats in terms of completion, behavioral engagement, knowledge gains, interest, and self-efficacy (RQ1).

\subsection{Interaction Modality in Dialogic Activities with AI}

Dialogue with AI can occur through text or voice, and the choice of modality may shape how students participate. \citet{clarkGroundingCommunication1991} explain that communication media impose different constraints on grounding, the process of establishing mutual understanding in communication. These include audibility, cotemporality, reviewability, and revisability, among others. These constraints shape how people produce, interpret, revisit, and clarify their contributions. Text and voice are therefore not interchangeable modality for the same dialogue and each shapes the interaction differently. Therefore, it is important to examine how interaction modality shapes students' engagement and learning. Although empirical research in educational settings remains limited, several studies have begun to compare these modalities. 

\citet{wambsganssDesigningConversationalEvaluation2022} have examined the effect of conversational interfaces in course evaluation, they found that both text- and voice-based conversational interfaces produced higher-quality responses and better user experiences than a static survey. In learning settings, however, comparisons between the two modalities have been mixed. One study found no significant differences in perceived value or quiz performance, while interviews showed that learners' preferences depended on their circumstances \citep{schlimbachEMBARKINGINTERACTIVELEARNING}. Another found that voice supported emotional and cognitive engagement, whereas text better supported autonomous learning and content review \citep{meleTalkingVsTyping2026}. These findings suggest that text and voice may affect how students engage, even when their learning outcomes are similar.

Research on language learning further shows that speaking and typing can create different cognitive and affective conditions. Both text and voice chat have been found to improve oral proficiency among learners of English, although reductions in anxiety were observed only for text chat in one study \citep{namaziandostSynchronousCMCText2022}. Conversational AI has also improved learners' willingness to communicate, perceived communicative competence, and speaking anxiety without necessarily improving speaking performance \citep{wangImpactDifferentConversational2024}. Conversely, a longer voice-based intervention produced gains in speaking fluency, interactional competence, and learner autonomy relative to peer-to-peer practice \citep{abdelradySpeakingAIImpact2026}. Although these studies concern language learning rather than learning academic content, they demonstrate that modality can shape learners' willingness and ability to participate independently of measured learning gains.

Prior work suggests not that one modality is superior, but that text and voice create different conditions for participation. Yet little is known about how these conditions shape multilingual students' engagement with academic content. We address this gap by examining how modality affects students' learning, engagement, and experiences in dialogic activities with AI (RQ2, RQ3).

\subsection{AI for Bilingual Learning Support}

For students learning academic content through a non-native language, the language of interaction adds another layer to these modality-specific demands. Cognitive Load Theory proposes that learning can be hindered when the demands of a task exceed the capacity of working memory \citep{swellerCognitiveLoadTheory2011, dejongCognitiveLoadTheory2010}. In EMI settings, students need to process unfamiliar concepts while simultaneously interpreting and producing academic English. \citet{rousselLearningSubjectContent2017a} has shown that learning in a foreign language puts students at a disadvantage because of the cognitive load. On the other hand, Researchers have shown that providing support in students' native languages can help mitigate these challenges \citep{alhazmiImpactBilingualVs2025a, rousselAdvantagesListeningAcademic2022, dejesusCognitiveLoadSupporting2025}.

AI could make such support more responsive and widely available. General-purpose tools including ChatGPT, QuillBot, and Grammarly have helped non-native English-speaking students manage linguistic and cognitive demands, understand course material, and personalize their learning \citep{tienExploringAImediatedLinguistic2025}. Studies in EMI and academic writing contexts likewise report improvements in understanding, motivation, writing efficiency, or writing quality, with some benefits being especially pronounced for non-native English speakers \citep{chroquiAIPoweredEMIClassroom2025, connellpenskyGenerativeAIsImpact2025, itoUseAIpoweredRewriting2023}. Students also use AI to move between their native language and English while reading and writing for academic purposes \citep{taylorTranslanguagingAIpoweredTools2026}. These practices show how AI can mediate access to English-language education, but they primarily position it as assistance used around an academic task rather than as a medium for the learning activity itself.

Multilingual AI tutors offer a path toward integrating language support into the learning process. In computing education, multilingual LLM tutors and programming chatbots have enabled students to seek explanations or produce prompts in their native languages \citep{kavuriFurtherStudyLeveraging2026, halaniExploringEffectivenessMultilingual2025, pratherBreakingProgrammingLanguage2025}. Researchers have examined AI teaching assistants \citep{chenIntegratingAITeaching2025}, chatbots \citep{zhouSupportingEnglishSpeaking2025, shashidharaVoiceBasedChatbotsEnglish2026, sharmaComuniqaExploringLarge2024, liuSingaKidsMultilingualMultimodal2025}, and digital avatars \citep{zhouResearchApplicationDigital2024, parkAITwinEnhancing2026} for conversation practice and language learning. However, research on tools designed to teach English answer a different question from tools that use multiple languages to help students learn academic content. Evidence from language practice therefore cannot by itself explain whether bilingual dialogue improves participation or learning in an EMI academic activity.

In summary, prior research has separately established the value of dialogic learning, identified distinct affordances of text and voice, and demonstrated ways that AI can support multilingual learners. These strands have not been studied together in an real-world EMI context. We address this intersection by comparing written responses with AI-based dialogic activities (RQ1), examining how text and voice dialogue operate in English-only and bilingual settings (RQ2), and investigating the factors that shape students' experiences and participation across these activities (RQ3).

\section{Study Methodology}

To address our research questions, we conducted an IRB-approved field study examining how the format and language of a learning activity shaped students' learning and engagement. 
Study participants were 305 undergraduate students in Karnataka, India, who were native Kannada speakers studying in English-medium institutions. In the study, all the participants learned the same content while the learning activity varied across the experimental conditions.

\subsection{Experimental Conditions}

We organized the experimental conditions along two primary dimensions: activity format and language. The activity-format dimension distinguished between \textit{Written Response} and \textit{Dialogic} activities. In the \textit{Written Response} format, students read a activity prompt and independently typed their response. In the \textit{Dialogic} format, students answered the same prompt by interacting with an AI tutor. Because interaction modality shapes how a dialogue unfolds, we implemented the dialogic format in two modalities: \textit{Text Dialogue}, in which students typed back and forth with the AI tutor, and \textit{Voice Dialogue}, in which they spoke with the AI tutor.

The language dimension distinguished between \textit{English} and \textit{Bilingual} activities. In the \textit{English} condition, the activity was conducted entirely in English. In the \textit{Bilingual} condition, students could use Kannada and English, with the interaction occurring primarily in Kannada.

After completing each learning activity, students received AI-generated feedback based on their written response or conversation with the AI tutor. The feedback was provided in the same language as the activity and included a written comment on their performance and areas for improvement. Students could then retry the activity. The timing of feedback varied by activity format. Students in the dialogic conditions received feedback immediately after completing the conversation, whereas those in the written response condition received feedback 12 hours after submitting their response. This delay was designed to reflect how students typically received feedback on conventional classroom assignments.  

In total, the study included six experimental conditions: written response and the two dialogic modalities (text dialogue and voice dialogue), each offered in English and bilingual versions. We used a between-subjects design, with each participant assigned to one condition for the duration of the study. Table~\ref{tab:enrollment-completion} reports enrollment and completion for each condition.

\begin{table*}[t]
\centering
\caption{Enrollment and completion across the Model Digital Inclusion Centers (MDICs) and college deployments, disaggregated by experimental condition.}
\label{tab:enrollment-completion}
\begin{tabular}{llccc}
\toprule
Activity/Modality & Language & \multicolumn{3}{c}{Enrolled (Completed)} \\
\cmidrule(lr){3-5}
 & & MDIC & College & Overall \\
\midrule
\multicolumn{5}{l}{\textit{Written Response}} \\
 & Bilingual & 26 (8) & 26 (26) & 52 (34) \\
 & English & 28 (13) & 25 (25) & 53 (38) \\
\cmidrule(lr){1-5}
\multicolumn{5}{l}{\textit{Dialogic}} \\
\multirow{2}{*}{\quad Text Dialogue} & Bilingual & 27 (6) & 26 (20) & 53 (26) \\
 & English & 27 (6) & 27 (24) & 54 (30) \\
\cmidrule(lr){1-5}
\multirow{2}{*}{\quad Voice Dialogue} & Bilingual & 30 (11) & 18 (14) & 48 (25) \\
 & English & 29 (7) & 16 (13) & 45 (20) \\
\midrule
\multicolumn{2}{l}{\textbf{Total}} & \textbf{167 (51)} & \textbf{138 (122)} & \textbf{305 (173)} \\
\bottomrule
\end{tabular}
\end{table*}

\subsection{Learning Module and Study Platform}

The study was conducted in partnership with the Kalvi Foundation\footnote{name anonymized for review.}, an education nonprofit organization in India, and embedded as a learning module within its ongoing AI literacy initiative to help students learn AI. We developed the module in collaboration with the Foundation's pedagogy team. It comprised four short bilingual instructional videos and corresponding learning activities. The videos reflected instructional practices common in participants' classrooms: technical and academic terms were retained in English, while explanations were provided in Kannada. After each video, students completed the corresponding activity in their assigned condition. The full learning module is described in Appendix~\ref{apx:learning-module}.

We delivered the study through a custom web-based learning management system (LMS) developed by the research team. The LMS was designed to resemble educational platforms familiar to students, such as Google Classroom, and was accessible on desktop and mobile devices. It provided a consistent structure across conditions while supporting the interfaces required by each activity format. Written response and text dialogue activities used web-based text interfaces, whereas voice dialogue activities used a real-time audio interface. The screenshots of the learning activity interfaces is provided in Appendix~\ref{app:interface-images}.

A unified AI backend supported the activities across conditions. Gemini 2.5 Flash handled all LLM-based components, including tutor responses in the dialogic conditions and post-activity feedback generation. For the voice-dialogue conditions, Pipecat\footnote{\url{https://github.com/pipecat-ai/pipecat}} orchestrated a real-time pipeline in which GPT-4o Transcribe converted students' speech to text, Gemini 2.5 Flash generated the tutor's response, and Cartesia Sonic 3 synthesized that response as speech. We selected and evaluated these models in collaboration with the implementation partner based on their output quality, latency, and language support, including their performance with Kannada. All dialogic conditions used the same core tutoring prompt -- only formatting instructions were adapted to the assigned language and modality.

Across conditions, we held constant the instructional videos, activity prompts, feedback rubric, and feedback generation prompt. Thus, the conditions differed in activity format, interaction modality, and language.

\subsection{Field Deployment and Study Procedure}

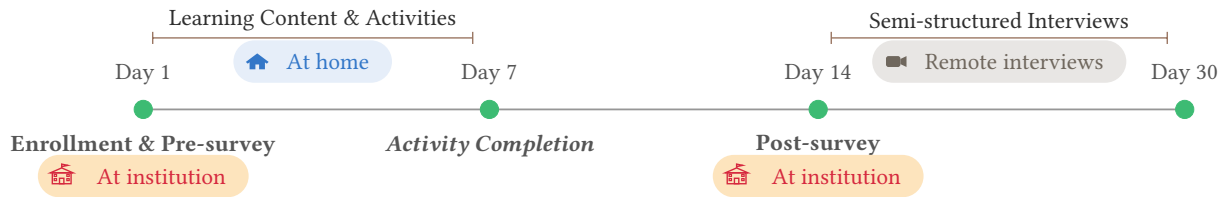
\begin{figure*}[t]
    \centering
    {\Large
    \resizebox{\textwidth}{!}{\usetikzlibrary{shapes.misc}

\definecolor{timelinegray}{HTML}{969696}
\definecolor{daygray}{HTML}{555555}
\definecolor{nodegreen}{HTML}{3DBB73}
\definecolor{bracketbrown}{HTML}{9A7662}
\definecolor{homeblue}{HTML}{3377C8}
\definecolor{homefill}{HTML}{E6EEF9}
\definecolor{institutionred}{HTML}{D6283E}
\definecolor{institutionfill}{HTML}{FDE4BD}
\definecolor{remotebrown}{HTML}{6F675D}
\definecolor{remotefill}{HTML}{E8E6E4}

\tikzset{
  day/.style={text=daygray},
  event/.style={font=\bfseries, text=daygray},
  phase/.style={text=black!82},
  pill/.style={rounded rectangle, minimum height=7mm, inner xsep=3.5mm,
  },
  timeline point/.style={circle, fill=nodegreen, draw=nodegreen,
    minimum size=3mm, inner sep=0pt},
  pics/house/.style={code={
    \fill[pic actions] (-.42,-.03)--(0,.34)--(.42,-.03)--(.33,-.03)--(.33,-.40)--
      (.08,-.40)--(.08,-.15)--(-.08,-.15)--(-.08,-.40)--(-.33,-.40)--(-.33,-.03)--cycle;
    \fill[pic actions] (-.49,-.02)--(0,.42)--(.49,-.02)--(.41,-.11)--(0,.25)--(-.41,-.11)--cycle;
  }},
  pics/building/.style={code={
    \draw[pic actions, line width=.65pt] (-.43,-.42) rectangle (.43,.18);
    \draw[pic actions, line width=.65pt] (-.52,.18)--(0,.45)--(.52,.18)--cycle;
    \draw[pic actions, line width=.65pt] (-.10,-.42) rectangle (.10,-.12);
    \foreach \x in {-.29,.29}{
      \draw[pic actions, line width=.55pt] (\x-.055,-.02) rectangle (\x+.055,.09);
      \draw[pic actions, line width=.55pt] (\x-.055,-.25) rectangle (\x+.055,-.14);
    }
    \draw[pic actions, line width=.65pt] (0,.45)--(0,.66)--(.28,.59)--(0,.52);
  }},
  pics/camera/.style={code={
    \fill[pic actions, rounded corners=.04cm] (-.46,-.30) rectangle (.20,.30);
    \fill[pic actions] (.20,.14)--(.48,.30)--(.48,-.30)--(.20,-.14)--cycle;
  }}
}

\begin{tikzpicture}[x=1.2cm,y=1cm]
  \draw[bracketbrown, line width=.65pt]
    (1.15,2.63)--(5.55,2.63)
    (1.15,2.53)--(1.15,2.73)
    (5.55,2.53)--(5.55,2.73);
  \node[phase] at (3.35,2.92) {Learning Content \& Activities};

  \draw[bracketbrown, line width=.65pt]
    (10.48,2.63)--(15.10,2.63)
    (10.48,2.53)--(10.48,2.73)
    (15.10,2.53)--(15.10,2.73);
  \node[phase] at (12.79,2.92) {Semi-structured Interviews};

  \node[pill, fill=homefill, text=homeblue] (home) at (3.35,2.24)
    {\hspace{5mm}At home};
  \pic[homeblue, scale=.32] at ([xshift=4mm]home.west) {house};

  \node[pill, fill=remotefill, text=remotebrown] (remote) at (12.79,2.24)
    {\hspace{5mm}Remote interviews};
  \pic[remotebrown, scale=.30] at ([xshift=4mm]remote.west) {camera};

  \draw[timelinegray, line width=.85pt] (1.02,1.45)--(15.34,1.45);

  \node[timeline point] at (1.02,1.45) {};
  \node[timeline point] at (5.78,1.45) {};
  \node[timeline point] at (10.30,1.45) {};
  \node[timeline point] at (15.34,1.45) {};

  \node[day, anchor=south] at (1.02,1.72) {Day 1};
  \node[day, anchor=south] at (5.78,1.72) {Day 7};
  \node[day, anchor=south] at (10.30,1.72) {Day 14};
  \node[day, anchor=south] at (15.34,1.72) {Day 30};

  \node[event, anchor=north] at (1.02,1.18) {Enrollment \& Pre-survey};
  \node[event, font=\bfseries\itshape, anchor=north] at (5.78,1.18)
    {Activity Completion};
  \node[event, anchor=north] at (10.30,1.18) {Post-survey};

  \node[pill, fill=institutionfill, text=institutionred] (institution-one) at (1.02,.35)
    {\hspace{6mm}At institution};
  \pic[institutionred, scale=.30] at ([xshift=4mm]institution-one.west) {building};

  \node[pill, fill=institutionfill, text=institutionred] (institution-two) at (10.30,.35)
    {\hspace{6mm}At institution};
  \pic[institutionred, scale=.30] at ([xshift=4mm]institution-two.west) {building};
\end{tikzpicture}}
    }
    \caption{Study implementation flow.}
    \Description{
Horizontal timeline of the study from Day 1 to Day 30. On Day 1, students enrolled and completed the pre-survey at their institution. They then completed the learning content and activities at home, with activity completion marked at Day 7. Students returned to their institution for the post-survey on Day 14. Semi-structured interviews were conducted remotely between approximately Day 14 and Day 30.
  }
    \label{fig:implementation-flowchart}
\end{figure*}

We conducted two sequential field deployments: first through 25 Model Digital Inclusion Centers (MDICs) operated by the Foundation, followed by a second deployment at an undergraduate college. The MDICs are community computer centers located in villages where students and other community members can access computers and learning programs. Across both deployments, students progressed through three stages: enrollment and a pre-study survey, completion of the learning content and activities, and a post-study survey one week later. Figure~\ref{fig:implementation-flowchart} summarizes this process.

In the first session, students provided informed consent and completed the pre-study survey. In the MDIC deployment, facilitators proctored the survey on center computers using Qualtrics. In the college deployment, the facilitator and professors proctored paper surveys because computers were not available for all students. For the paper administration, we prepared two versions of the survey, with questions in different orders, to reduce the possibility of copying. The research team subsequently digitized the completed paper responses.

After completing the survey, students enrolled in the LMS and were randomly assigned to one of the six experimental conditions, which they retained throughout the study. Students then completed a sample activity to familiarize themselves with the LMS and their assigned activity format.

Students subsequently completed the learning content and activities primarily from home using their smartphones. Facilitators remained available to answer questions and provide clarification, including through WhatsApp groups. Participants completed the post-study survey one week after their final activity. At each deployment, the post-study survey used the same administration mode and proctoring procedure as the pre-study survey. Students received a certificate of participation from the implementation partner upon completing the study.

The MDIC deployment enrolled 167 undergraduate students aged 18--23, of whom 51 completed the study. The lower completion rate was partly attributable to an unexpected interruption two weeks into the program, when the MDICs shut down permanently because of unforeseen funding cuts affecting the implementation partner. The subsequent college deployment enrolled 138 undergraduate students aged 18--23, of whom 122 completed the study. In total, 305 students enrolled and 173 completed the study. Enrollment and completion by condition are reported in Table~\ref{tab:enrollment-completion}.

\subsection{Data Collection and Analysis}

We drew on three complementary data sources --- pre- and post-study surveys, LMS interaction logs, and semi-structured interviews --- to investigate learning outcomes, examine patterns of engagement, and understand students' experiences with the learning activities. Unless otherwise stated, our analyses focused on the 173 students who completed the study. The discourse analysis was the exception --- it included all students who generated conversational data, regardless of study completion. We included all students in the discourse analysis to identify conversational patterns that might help explain study non-completion.

\subsubheading{Pre- and Post-study Surveys} \label{section:surveys}

The surveys measured one cognitive outcome (knowledge of the topic) and two affective outcomes (interest in the topic and self-efficacy related to the topic). The post-study survey additionally measured students' preference for the study activity relative to their usual learning activities. The full survey instruments are provided in Appendix~\ref{apx:surveys}.

\parabold{Knowledge Assessment}
We measured knowledge of the course concepts using 15 multiple-choice and open-ended questions, such as \textit{Do AI systems always make accurate predictions? Why or why not?} Each question was worth one point, for a maximum score of 15. The research team developed the questions with the implementation partner's pedagogy team. The same items appeared in both surveys, allowing us to assess change over time and compare post-study outcomes across experimental conditions.

Open-ended responses were scored using a rubric developed by the research team. A subset of 236 open-ended responses was first graded by the research team. We then scaled grading to the full sample using GPT-5.4-mini with the same rubric. LLM grading showed strong reliability against human ratings (quadratic-weighted $\kappa = .85$, ICC = .85), and the research team manually verified all LLM-assisted grades. Overall, the 15-item knowledge assessment showed strong internal consistency in our sample ($\alpha = .79$ pre; $\alpha = .80$ post).

\parabold{Interest}
We measured students' interest in learning about AI to examine whether it differed across experimental conditions. We used the five-item validated scale developed by \citet{hornbergerMultinationalAssessmentAI2025}. Students responded on a five-point Likert scale ranging from strongly disagree to strongly agree. The scale showed strong internal consistency in our data ($\alpha = .72$ pre; $\alpha = .80$ post).

\parabold{Self-efficacy}
We measured students' confidence in understanding AI to examine whether it differed across experimental conditions. We used the eight-item validated scale developed by \citet{hornbergerMultinationalAssessmentAI2025}. 
Students responded on a five-point Likert scale. The scale showed strong internal consistency in our data ($\alpha = .82$ pre; $\alpha = .83$ post).

\parabold{Activity Preference}
In the post-study survey, students compared the study activity with their usual learning activities, such as completing written homework assignments. They responded on a five-point scale ranging from strongly preferring the study activity to strongly preferring their usual activities. This measure captured activity preference independently of learning outcomes.

\parabold{Survey Analysis}
For knowledge, interest, and self-efficacy, we fitted regression models predicting each post-study score from the corresponding pre-study score, activity condition, language condition, their interaction, and deployment site. We report model-adjusted means with HC3 robust standard errors. Because activity preference was measured only after the study, we analyzed it separately across experimental conditions. We report other statistical tests alongside their corresponding results in the Findings.

\subsubheading{LMS Interaction Logs} \label{section:log-analysis}

The LMS logged activity start and completion timestamps, responses submitted in the written response condition, keystroke data, and conversation logs from the text- and voice dialogue conditions. Voice conversations were transcribed before analysis. We used these data to examine both behavioral engagement and the content of students' interactions with the AI tutor.

\parabold{Quantitative Engagement Analysis}
We quantified engagement using time on task, total words produced by students, and the number of conversation turns in dialogic activities. Time on task was calculated as the elapsed time between the logged start and completion of an activity; inactive intervals within this period were not removed. Total word count captured the amount of language students produced, and turn count captured the extent of back-and-forth interaction with the AI tutor. We report the statistical tests alongside the corresponding results.

\textit{AI-assisted Responses.}
Because some students used external AI assistance when producing text responses, we examined the written response and text dialogue conditions for evidence of AI-generated content. The second author flagged suspected responses, and the first author verified them. The primary indicator was unusually long response length compared to the typical response length of other responses in the same activity, supplemented by frequent emoji use and keystroke evidence that text had been pasted rather than typed.

These indicators could be applied only to typed responses. We found no comparable evidence in voice conversations, where students would have needed to read generated text aloud. Voice turns averaged 3.4 words, and even the most talkative students averaged 9.4 words per turn, whereas typed responses flagged as potentially AI-assisted averaged 108 words. We therefore treat the absence of flagged voice responses as evidence that this shortcut was uncommon, not proof that it never occurred.

We flagged 9.3\% of written responses (8.6\% in English; 10.0\% in bilingual) and 7.4\% of text dialogue responses (6.8\% in English; 8.1\% in bilingual). For word-count analyses, we subtracted words in flagged responses from each student's total so that the measure better reflected student-produced language. We retained flagged responses in turn counts because each response still represented a student turn. We report word-count comparisons both with and without this adjustment.

\parabold{Discourse Analysis} \label{section:conversation-analysis}
We analyzed conversations from the dialogic conditions to examine how students engaged with the AI tutor and how their interactions unfolded. The full corpus contained 1,806 conversations from 167 students. 
We excluded 202 \textit{sample activity} conversations used to familiarize students with the system, leaving 1,604 conversations for analysis.

The research team developed a coding scheme through deductive and inductive coding. We drew initial codes from prior discourse-analysis work in educational contexts \cite{stolckeDialogueActModeling2000, sureshTalkMovesDataset2022, zhouTutorMoveTaxonomy2026} and refined them based on patterns in the corpus. The final codebook contained four student-level codes, four AI tutor-level codes, and one conversation-level code. Student-level codes captured demonstrated understanding, emerging understanding, confusion, and unrelated responses; confusion and unrelated responses were characterized as student friction. AI tutor-level codes captured content delivery, elicitation of student thinking, affective support, and deviation from the learning goal; deviation from the learning goal was characterized as agent friction. These turn-level codes were not mutually exclusive. The conversation-level code captured whether the interaction reached resolution: whether the student arrived at an adequate response to the activity at the end of the conversation without ending the conversation midway. The complete codebook is provided in Appendix~\ref{apx:codebook}.

To validate automated annotation, the research team manually labeled 100 conversations randomly sampled across the four dialogic experimental groups (25 per group). We used these annotations as gold labels to compare Gemini 3.7 Flash and Claude Sonnet 5. Gemini showed higher mean agreement with the human labels across the nine codes ($\kappa = .60$ versus $.54$). Agreement was moderate or substantial for eight codes. For quantitative comparisons, we retained only the three codes that reached substantial agreement: conversation resolution ($\kappa = .85$), demonstrated understanding ($\kappa = .65$), and student friction ($\kappa = .65$).

\subsubheading{Semi-structured Interviews}
We conducted semi-structured interviews with 16 students after course completion to understand the students' experiences with the different learning activities. We sampled across experimental conditions and levels of engagement to capture a range of experiences. Table~\ref{tab:interview-allocation} shows the distribution of interview participants.

\begin{table}[t]
\centering
\caption{Allocation of the 16 interview participants by activity condition and language.}
\label{tab:interview-allocation}
\begin{tabular}{lccc}
\toprule
Activity Condition & Bilingual & English & Total \\
\midrule
Voice Dialogue & 3 & 3 & 6 \\
Text Dialogue & 3 & 3 & 6 \\
Written Response & 2 & 2 & 4 \\
\midrule
Total & 8 & 8 & 16 \\
\bottomrule
\end{tabular}
\end{table}

Interviews were conducted remotely through Google Meet, Zoom, or WhatsApp Audio Call, depending on students' convenience. They were conducted in English and Kannada, with a facilitator from the implementation partner translating in real time when necessary. The protocol covered students' educational backgrounds, course experiences and learning, experiences with the activities, comparisons with usual coursework, and uses of AI outside the course. The full protocol is provided in Appendix~\ref{apx:interview-protocol}.

We analyzed the transcripts using thematic analysis \cite{feredayDemonstratingRigorUsing2006}. The research team coded the interviews in ATLAS.ti\footnote{\url{https://atlasti.com/}}, beginning inductively with codes developed from the data. We iteratively reviewed, refined, and merged the codes into a final codebook of five themes and 27 codes. The themes were Patterns of AI Use, Experience of the AI Tutor, Language Preferences and Trade-offs, Modality Preferences and Trade-offs, and Perceived Effects of AI on Learning. The full codebook, including code counts, is provided in Appendix~\ref{apx:interview-codebook}.



\section{Findings}

Our findings show how AI-based dialogic activities 
shaped students’ participation and learning experiences. We begin by comparing written and dialogic activities in terms of participation, completion, and students’ preferences. We then examine differences within the dialogic activities, drawing on activity records, discourse analysis, and interviews to understand how text, voice, and language created different opportunities and challenges for engagement. Next, we consider how these activity formats and language conditions related to knowledge gains, interest, and self-efficacy. Finally, we outline students’ broader aspirations around language and their existing uses of AI, which ranged from supporting their reasoning to over-reliance.

\subsection{Impact on Engagement and Completion}\label{sec:impact-engagement-completion}

We examined both the quantity and quality of students' engagement. We measured the quantity of engagement through the number of words students produced, the time they spent on the activity, and, for the dialogic activities, the number of conversational turns. We complemented these measures with a discourse analysis of the dialogic interactions to examine the quality of students' participation and with interviews to understand how students experienced the different activity formats and language conditions.

\begin{figure*}[t]
\centering
\includegraphics[width=\textwidth]{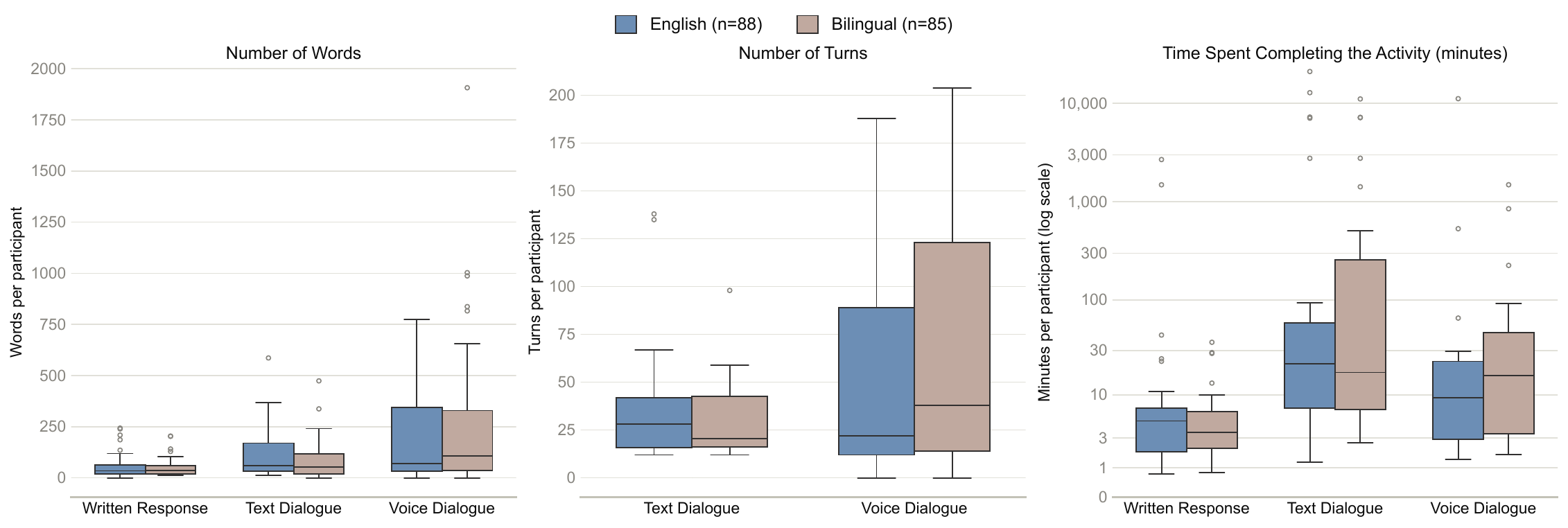}
\caption{Engagement by activity format and language among the $N = 173$ participants who completed the activity (English $n = 88$, Bilingual $n = 85$): (a) words produced per participant, after subtracting words flagged as AI-generated; (b) conversational turns per participant; and (c) participation time in minutes, shown on a $\log(1+x)$-spaced axis. Panel (c) uses this scale because the largest values may include idle time from participants who left the activity open on their device.}
\Description{
Three sets of box plots compare engagement for English and bilingual participants. Panel A shows words produced: written responses have the lowest word counts, text dialogue is higher, and voice dialogue is highest and most variable. Panel B shows conversational turns for text and voice dialogue; their typical turn counts are similar, but voice dialogue has substantially greater variation, with some participants contributing very few turns and others more than 200. Panel C shows time spent on a logarithmic scale: participants spend considerably more time in both text and voice dialogue than in written response. English and bilingual distributions generally overlap, although bilingual voice participants tend toward higher engagement. Several extreme time values may represent activities left open while participants were inactive.
}
\label{fig}
\end{figure*}

\subsubheading{Dialogic activities elicited higher engagement}\label{subsubsec:higher-engagement}

Students in the voice dialogue condition produced responses with a median length of 102 words, compared with 58.5 words in text dialogue condition and 34 words in written response condition. 
Response length differed significantly across the three activity formats (Kruskal--Wallis $H = 14.42$, $p < .001$). 
Both dialogue conditions elicited longer responses than the written response condition (voice dialogue $p < .001$, text dialogue $p = .019$). Although responses were longer in the voice condition than in the text condition, this difference was not statistically significant ($p = .075$). Figure~\ref{fig}(a) shows the full distributions.

These response length counts excluded responses flagged as AI-generated, as described in \S\ref{section:conversation-analysis}, so that we compare only the words students produced themselves. Including AI-generated text inflated word counts in the text conditions (written response and text dialogue) since AI-generated responses were typically very long. However, the overall pattern was unchanged: whether AI-generated text was included or excluded, both dialogue conditions produced more words than the written response condition.



The dialogue-based activities also sustained students’ participation for much longer than the written response activity. Students spent a median of 19 minutes in text dialogue and 14.6 minutes in voice dialogue, compared with only 3.8 minutes in written response. Participation time differed significantly across the three formats (Kruskal--Wallis $H = 37.49$, $p < .001$), with students spending significantly longer in both text and voice dialogue than in written response ($p < .001$ for both comparisons). Although the median participation time was longer for text than voice dialogue, the difference was not statistically significant ($p = .087$).

Our discourse analysis revealed that conversations may have lasted longer because the tutor continued to support students when they did not know an answer. Instead of ending the exchange, the tutor reassured them, provided hints and feedback, and asked follow-up questions. This support helped students continue responding and develop their understanding over several conversational turns. For example, in one interaction, a student repeatedly said that they don't know the answer about two minutes into the conversation, the tutor reassured the student, explained part of the concept, and asked a follow-up question: \verbinline{\small If you don't know, it's okay, I will help....What should humans do to give meaning to those cluster groups?}
The student then continued, and the conversation lasted about five minutes.


\subsubheading{Voice produced more heterogeneous patterns of engagement}\label{subsubsec:voice-interaction}
Voice and text dialogue did not differ significantly in the typical number of conversational turns. Students completed a median of 33 conversational turns in voice dialogue and 24.5 turns in text dialogue, but this difference was not statistically significant ($p = .456$). The more notable difference was in how much participation varied across students. In the voice condition, turn counts ranged from 0 to 204, with an interquartile range of 13 to 92. In contrast, in the text condition, they ranged from 12 to 138, with a much narrower interquartile range of 15.8 to 42.3. These distributions show that voice dialogue produced highly uneven participation:  some students sustained long conversations, while others contributed very little or did not participate at all. 
In contrast, participation in text dialogue was more consistent across students. This difference in dispersion was significant (Brown--Forsythe $F(1,99) = 16.44$, $p < .001$), and it persisted after log transformation, suggesting that it was not driven only by a few unusually long conversations. Our interviews revealed reasons why voice modality produced such different experiences. For some, speaking reduced the effort needed to sustain the interaction. Rather than having to formulate and type each response, they could speak more freely and maintain a conversational flow. As P9 explained:

\begin{quote}
\textit{``When I am texting, I find it tiring to keep the conversation going because I am already using my brain power to think and text, and then again I have to think, what next, what next? But when it comes to speaking, I can speak very easily because there is a flow to it.''} (P9, Voice, Bilingual)
\end{quote}

For others, however, voice introduced additional demands because they had to listen, remember, and respond in real time. P8 preferred text because it allowed them to revisit earlier exchanges:

\begin{quote}
\textit{``When I am having a conversation through text chat, I can easily refer back to the previous exchanges, especially if I forget or lose track of what we were talking about. But if I am having a voice-to-voice conversation, sometimes I forget what I wanted to ask.''} (P8, Voice, Bilingual)
\end{quote}

The pacing and the naturalness of the conversation also shaped students' engagement. 
In our interviews, students appreciated AI tutor’s voice and expressiveness, which made the interaction feel more like a conversation with another person. Yet this human-like quality did not always make the conversation seamless. Students reported that the AI tutor in voice dialogue condition often did not pace the interaction effectively, a pattern we also observed in our discourse analysis of the audio recordings. When students paused to formulate an answer, the tutor often began responding immediately. These pauses were treated as the end of a turn, leading to interruptions and overlapping speech and putting pressure on students to respond quickly.


Voice dialogue therefore did not uniformly reduce the effort of participation. It made conversation easier and more fluid for some students, while creating additional cognitive and interactional demands for others. These contrasting experiences help explain why participation varied much more widely in voice dialogue than in text dialogue.

These demands became more pronounced when students also had to formulate their responses in English. Students with limited fluency in spoken English often needed additional time to construct sentences and express their ideas. Yet the tutor frequently interpreted these pauses as the end of a turn and began responding before the student had finished. This led to repeated interruptions and left students with little opportunity to complete or develop their answers. The combined pressure of speaking in English, responding quickly, and coordinating turns made these interactions difficult to sustain, as illustrated in the conversation below from the voice dialog condition.

\begin{quote}
\small

\noindent\textbf{AI Tutor:} \verbinline{\small No worries at all. Think about it this way. 
The AI is looking at all the user data and finding patterns to put similar 
users together. What's that process?} \textit{[Student interrupts]}

\smallskip
\noindent\textbf{Student:} It is quantitative data. 
\textit{[AI Tutor interrupts]}

\smallskip
\noindent\textbf{AI Tutor:} \verbinline{\small That's a good observation about the type of data. 
What's the AI doing with that quantitative data to create groups?}

\smallskip
\noindent\textbf{Student:} It is quantitative data used to create a grouping 
systematically. \textit{[AI Tutor interrupts]}

\smallskip
\noindent\textbf{AI Tutor:} \verbinline{\small You've got the systematic.} 
\textit{[Student interrupts]}
\end{quote}


\subsubheading{Language shaped participation differently in voice and text dialogue}\label{subsubsec:language-intricacies}
While participation in voice dialogue varied widely across students, we found that language shaped how students engaged with the AI tutor. In the voice dialogue condition, students assigned to the bilingual voice condition showed consistently higher levels of participation than those who were assigned to English-only condition.  Compared with participants in the English voice condition, those in the bilingual voice condition had higher median word counts (108 vs. 69.5), turn counts (38 vs. 22), and participation time (16.3 vs. 9.2 minutes). These differences were not statistically significant, so they do not provide sufficient evidence of an overall difference in the amount of participation between the two groups. However, the fact that all three measures showed the same directional pattern suggested a possible relationship between language condition and participation. We therefore examined the conversations more closely, finding a clearer difference not in how much students participated, but in how language shaped their experience of the voice interaction. 

\parabold{Bilingual voice dialogue showed fewer communication breakdowns}
Our discourse analysis showed that a larger proportion of student responses demonstrated topical understanding in bilingual voice dialogue than in English voice dialogue (43.9\% vs. 20.5\% of turns; student-clustered GEE, OR = 2.29, $p = .001$). An opposite pattern emerged for articulation difficulty: 44.3\% of student responses in English voice dialogue showed difficulty formulating a response, compared with 17.5\% in bilingual voice dialogue (student-clustered GEE, OR = 0.52, $p = .018$).
These difficulties were also reflected in whether students sustained the conversation. Students abandoned 86.7\% of English voice conversations before completion, compared with 65.4\% of bilingual voice conversations (student-clustered GEE, OR = 0.32, $p = .004$).
In the English voice dialogue, students sometimes struggled to formulate a response, said \textit{``I don't know,''} or switched to Kannada. When the tutor continued to ask for a response in English, the exchange could stall or end. 

Even when students could continue the conversation, limited English speaking skills often led to difficulties in expressing learned concepts effectively. The conversation below from 
illustrates how difficulty expressing an idea in English could lead to misunderstanding. The student was trying to explain that Google provides answers to questions easily. The tutor, however, interpreted the comment as a request for a direct answer and explained that its role was to support learning rather than provide answers. Even after the student tried to clarify, the tutor repeated the same explanation rather than recognizing what the student meant.
\begin{quote}
\small
\noindent\textbf{AI Tutor:} \verbinline{\small I'm not sure I understand. Can you tell me in English how you use Google?}

\smallskip
\noindent\textbf{Student:} It used to tell me the easily answer questions.

\smallskip
\noindent\textbf{AI Tutor:} \verbinline{\small I'm not here to give you the answers directly. I'm here to help you think through it. Let's take Google as an example. When you use Google, what do you do? And what information might Google learn from that?}

\smallskip
\noindent\textbf{Student:} I tell Google to question. Then, it tell me the easily answer.

\smallskip
\noindent\textbf{AI Tutor:} \verbinline{\small I'm here to help you learn, not just give you the answers. What happens when you search for something on Google?}

\end{quote}


Similar breakdowns occurred more frequently in English voice conversations. At times, the tutor failed to recognize students’ speech or misinterpreted their responses. At other times, students repeatedly asked the tutor to repeat itself because they had difficulty understanding it. The effort required to identify and repair these misunderstandings disrupted the exchange and made the conversation harder to sustain.
In contrast, 
students expressed their ideas and understood the tutor more readily in bilingual voice dialogue. Ability to use AI tutor in native language allowed the conversation to continue even when the student struggled to understand the topic or made conceptual errors. Rather than stalling, the interaction enabled the student to respond to the tutor’s guidance, revise their understanding, and develop their ideas over multiple turns.


\parabold{Text supported English formulation, while voice enabled more natural Kannada conversation}
The language differences observed in voice dialogue were not evident in text dialogue. Our discourse analysis showed that student responses demonstrated understanding of learned concepts at similar rates in the bilingual and English text conditions (37.6\% vs. 32.2\% of turns; student-clustered GEE, $p = .34$). Moreover, English text conversations were not abandoned significantly more often than bilingual text conversations (62.4\% vs. 49.1\%; student-clustered GEE, $p = .18$). In our interviews, students explained that text made it easier to communicate in English by giving them more time and control over their responses. Unlike in voice dialogue, students could inspect what they had written and correct mistakes before responding. This additional time was particularly useful for students who understood English but were less comfortable speaking it fluently. As P10 explained:



\begin{quote}
\textit{``Some students like me have a spoken English problem. We can understand English well but it is difficult for us to speak fluently. If we do text messages, we get more time to frame our sentences properly in English before responding.''} (P10, Text Dialogue, English)
\end{quote}

The benefits of text dialogue, however, differed by language. While typing gave students time to formulate their responses in English, students did not need the same support to express themselves in Kannada. Because they were already proficient Kannada speakers, they found it easier to respond by speaking. Voice dialogue also felt more like a casual conversation, allowing students to use Kannada and move between languages in ways that were already familiar to them.

Students were also accustomed to typing Kannada using Latin script in informal chats, and we observed this practice in their responses. However, they had little prior experience using such bilingual writing in learning activities. Their written assignments had generally required them to respond either in English or entirely in Kannada script, rather than using Kannada written in Latin script or moving between languages. Thus, although Latin-script Kannada was familiar from informal communication, using it to complete a learning activity was less familiar.

Typing in Kannada script introduced an additional difficulty. Kannada uses a larger and more complex set of characters than English, including combinations of consonants, vowels, and diacritics that can require multiple keystrokes to enter. Kannada script appeared in 2.64\% of student text entries (0.49\% in bilingual written responses and 3.62\% in bilingual text dialogue turns). The other entries used predominantly Latin characters and could contain English, Kannada written in Latin script, or a mixture of the two. Latin script therefore allowed students to avoid some of the difficulties of the Kannada keyboard. Still, when voice was available, students generally found it easier to speak Kannada than to type it in either script. P10 described this contrast:

\begin{quote}
\textit{``In English I would prefer typing, but in Kannada I would prefer talking because in keyboard, to type in Kannada is difficult. English is easier to type than Kannada.''} (P10, Text Dialogue, English)
\end{quote}


\subsubheading{Dialogic activities were preferred but completed less often}\label{subsubsec:dialogic-less-completion}
Although students preferred both dialogue formats over written response, they were less likely to complete the dialogic activities. Across all 305 enrolled students, 68.6\% completed written response compared with 52.3\% in text dialogue and 48.4\% in voice dialogue. Completion rates differed significantly across the three activity formats (Pearson $\chi^{2}(2) = 9.48$, $p = .009$). 

In contrast, overall completion did not differ by language condition. Completion rates were similar in the English and bilingual conditions (57.9\% vs. 55.6\%, $p = .77$), suggesting that activity format, rather than language alone, was more closely associated with whether students completed the activity.


Completion also differed between the two deployments. The first deployment ended early because of a funding cut, limiting the time available for students to complete the activity. To separate the effect of this disruption from differences across activity formats, we examined the college deployment independently, where students had the full intended period to participate.
The same pattern remained, with completion rates differing significantly across activity formats: 100\% of students completed the written activity, compared with 83.0\% in text dialogue and 79.4\% in voice dialogue (Pearson $\chi^{2}(2) = 10.87$, $p = .004$). The lower completion of the dialogic activities therefore cannot be explained only by the early end of the first deployment. Within the first deployment, completion was low in all conditions, and differences between activity formats were not statistically significant ($p = .17$). Table~\ref{tab:enrollment-completion} provides the breakdown by condition and deployment site.

Among students who completed the activity and responded to the post-survey, the dialogic activities were evaluated more favorably than written response. When comparing the study activity with their usual learning activities (for example, writing answers as homework), 68\% of students in the dialogue conditions preferred the study activity, while 17\% preferred their usual activities.
In contrast, in written response condition, 14\% preferred the study activity and 47\% preferred their usual learning activities. Preference differed significantly across activity formats (Kruskal--Wallis $H = 39.53$, $p < .001$), and both dialogic formats differed from written response ($p < .001$ in each case).

Preference did not vary systematically by language. In a regression that included activity format, language, and deployment site, language did not independently predict preference ($p = .85$). There was also no clear evidence that the effect of activity format differed by language ($p = .20$). The main difference was therefore between written response and the two dialogic activities, not between English and bilingual interaction or between text and voice dialogue. P15 explained this preference by contrasting the support available during dialogue with the one-way nature of a typical written assignment:

\begin{quote}
\textit{``Usually, in assignments, we have to write and complete everything on our own. But here, when we talk with the bot, we gain more knowledge and develop better understanding of the topic. Even if I don't know much about something, the bot provides additional information and immediate feedback, which helps me learn more. In written assignments, information is more one-way.''} (P15, Voice Dialogue, Bilingual)
\end{quote}

Overall, students who completed the dialogic activities spent more time on them, contributed more, and preferred them to written response. However, fewer students completed the dialogic activities. Written response showed the opposite pattern: more students completed it, but those who did spent less time and contributed less.

\subsection{Learning Outcomes and Students' Experiences with AI}\label{subsec:learning-with-ai}

\begin{figure*}[t]
  \centering
  \includegraphics[width=\textwidth]{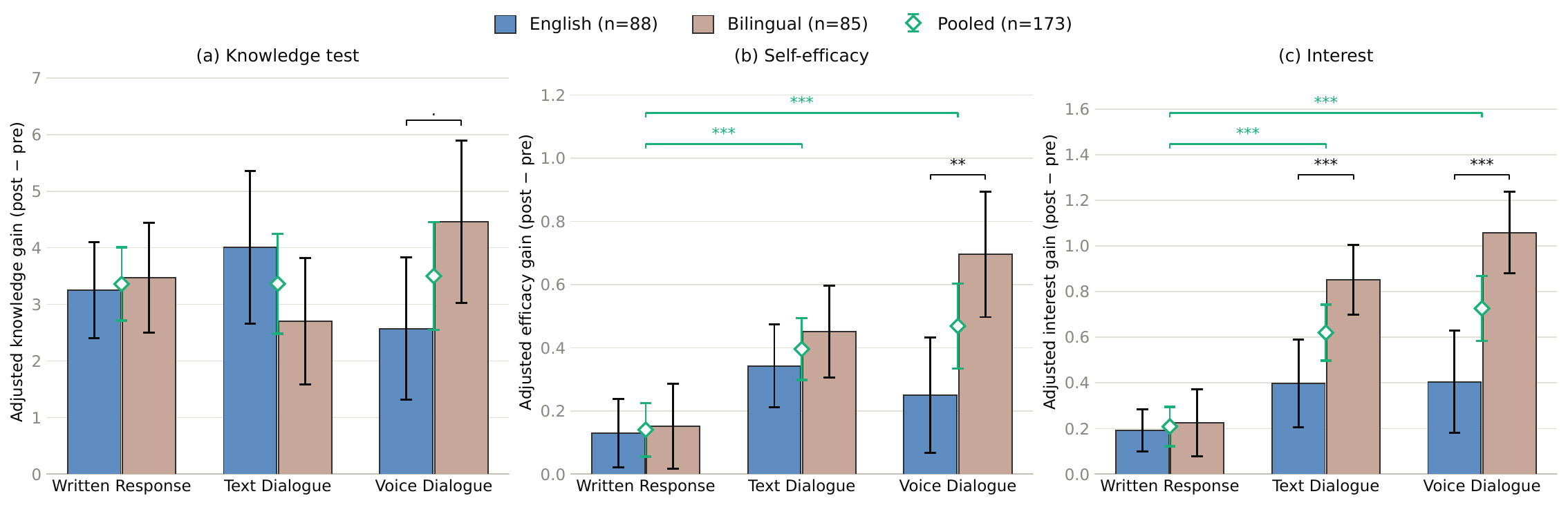}
\caption{Model-adjusted pre-to-post gains by activity format and language. Bars show estimated marginal means from OLS regressions of each post-activity outcome on its
pre-activity counterpart, activity format $\times$ language condition, and deployment site ($N = 173$, HC3 robust standard errors), averaged over site weighted by its share of the sample. Estimates are expressed as gains relative to the sample's average pre-activity score: (a) knowledge, relative to a baseline of $5.45$; (b) self-efficacy, relative to a baseline of $3.13$; and (c) interest, relative to a baseline of $3.34$. Error bars show 95\% confidence intervals. Black brackets test Bilingual vs.\ English within each format; teal brackets test every pair of formats, pooled across language. Both are shown only where significant; a comparison without a bracket is not significant ($p \geq 0.10$). $^{***}p<0.001$, $^{**}p<0.01$, $^{*}p<0.05$,
$^{.}p<0.10$.}
\Description{Three bar-chart panels compare adjusted pre-to-post gains for written response, text dialogue, and voice dialogue, separately for English and bilingual participants; markers also show estimates pooled across language, and error bars show 95\% confidence intervals. Panel A shows knowledge gains in every condition, with no reliable differences among activity formats or languages; bilingual voice has the largest estimated gain, but its advantage over English voice is only marginal. Panel B shows self-efficacy: both dialogue formats have significantly larger pooled gains than written response, and bilingual voice has a significantly larger gain than English voice. Panel C shows interest: both dialogue formats again have significantly larger pooled gains than written response, while bilingual participants have significantly larger gains than English participants in both text and voice dialogue.}

  \label{fig:adjusted-means-panel}
\end{figure*}

Before and after the activity, students completed a 15-point knowledge test and five-point measures of self-efficacy and interest.
To compare these three outcomes across conditions, we used regression models that predicted students’ post-activity scores based on their pre-activity scores, activity format, language condition, and deployment site. This allowed us to compare the conditions while accounting for differences in students’ starting scores. Figure~\ref{fig:adjusted-means-panel} presents the model-adjusted pre-to-post gains.

\subsubheading{Knowledge increased across all conditions, but gains were similar}\label{subsubsec:knowledge-gain}

Students demonstrated substantial learning regardless of activity format or language condition. On the 15-point knowledge test, mean scores increased from 5.45 before the learning activity to 8.89 afterward (Wilcoxon $p < .001$). Knowledge scores also increased significantly within each of the three activity formats ($p < .001$ in all three cases), indicating that students learned through written response, text dialogue, and voice dialogue. 
However, we found no reliable evidence that one format or language produced greater knowledge gains than another. The two dialogic formats together differed from written response by only 0.07 points ($p = .88$), while voice and text dialogue differed by 0.16 points ($p = .81$). Bilingual participants also did not reliably outperform English participants overall (0.27 points, $p = .59$). Thus, although students preferred the dialogic activities and spent more time on them, these differences did not translate into reliably greater short-term knowledge gains.

Bilingual voice dialogue produced the highest adjusted knowledge score (9.91), although knowledge scores did not differ significantly across the six experimental groups. 
Within voice dialogue, bilingual voice scored 1.89 points higher than English voice ($p = .052$). In text dialogue, English scored 1.31 points higher than bilingual text dialogue ($p = .14$). Given the small number of students in each group (approximately 20--30), these results do not provide sufficient evidence that language affected knowledge gains. However, the bilingual voice condition had the highest adjusted knowledge score, and its estimated advantage over English voice dialogue was suggestive but inconclusive (1.89 points, $p = .052$). Further research with larger samples is needed to determine whether these observed differences represent reliable effects on learning.

\subsubheading{Dialogic activities increased interest and self-efficacy}\label{subsubsec:increased-interest}
Although knowledge gains were similar across activity formats, the dialogic activities produced stronger gains in students’ interest and self-efficacy. On the five-point interest scale, the two dialogic formats together scored 0.47 points higher than written response ($p < .001$). This advantage was evident for both text dialogue (0.42 points, $p < .001$) and voice dialogue (0.52 points, $p < .001$).

The difference was smaller for self-efficacy but followed the same pattern. Together, the dialogic formats scored 0.29 points higher than written response ($p < .001$), with significant differences for both text dialogue (0.26 points, $p < .001$) and voice dialogue (0.33 points, $p < .001$). These results show that  dialogic formats left students feeling more interested in the topic and more confident in their ability to understand it.


\parabold{Responsive guidance helped students continue reasoning}
Our interaction logs, interviews, and discourse analysis help explain how the dialogic activities supported students’ interest and self-efficacy. Across the conversations, the tutor responded to students' answers with immediate feedback, follow-up questions, examples, and hints. The conversation below shows how the bilingual voice tutor built on a student's brief responses rather than simply marking them as correct or incorrect.

\begin{quote}
\small

\noindent\textbf{AI Tutor:} \verbinline{\small In the digital services you have chosen, can you explain how data is used?}

\smallskip
\noindent\textbf{Student:} Analyzing the information.

\smallskip
\noindent\textbf{AI Tutor:} \verbinline{\small Yes, you are thinking in the right direction. Analyzing the information is a key part of data usage. But how is this analysis used to make specific decisions or provide advice? For example, think about the videos you watch on YouTube. How does data analysis work there?}

\smallskip
\noindent\textbf{Student:} Collecting user data.

\smallskip
\noindent\textbf{AI Tutor:} \verbinline{\small Definitely, that is also very important. Collecting data is the first step. But after collecting it, what do they do with that data?}

\smallskip
\noindent\textbf{Student:} Hmm, okay.

\smallskip
\noindent\textbf{AI Tutor:} \verbinline{\small Take time to think. This is a slightly complex subject, but you will definitely understand. Okay, I'll give you a clue. As you said, data is collected first. Then you said, ``analyzing the information.'' What do they understand from this analysis? What are they trying to find out?}

\end{quote}


As the conversation shows, the student initially offered brief, partially developed responses. Rather than rejecting these responses, the tutor acknowledged what was relevant and asked follow-up questions to encourage further explanation. It also introduced familiar examples to make the question more concrete. When the student hesitated, the tutor provided reassurance and a hint based on the student’s earlier responses. The student then offered a more specific answer, illustrating how responsive guidance could help sustain the exchange and support students in developing their reasoning over multiple turns.

\parabold{Encouragement helped students feel more confident}
Alongside providing guidance, the tutor frequently encouraged students when they hesitated, made mistakes, or did not know an answer in both text and voice dialogue conditions. Our interviews showed that reassurance from the tutor made uncertainty feel more manageable for many students. P11 explicitly connected the tutor's encouragement to an increased sense of confidence:

\begin{quote}
\textit{``Sometimes even when I didn't know the answer, it (the bot) was saying, `You don't worry.' So, it brought me confidence.''} (P11, Text Dialogue, English)
\end{quote}

Similar encouragement appeared in the conversation shared earlier. 
When the student hesitated, the tutor encouraged them to take time to think and reassured them that they would understand the topic before offering a clue. Such responses encouraged the students to continue despite their uncertainty.

\parabold{Students experienced AI as a less judgmental space}
Students also felt more comfortable asking the AI tutor questions because they did not expect it to judge them for not knowing an answer or understanding the material. This was particularly important for students who felt shy about revealing uncertainty or asking what they worried might be a basic question in front of teachers or peers. As P9 explained:

\begin{quote}
\textit{``I see a few of my friends who are a bit shy to ask questions to teachers. But I feel that this bot would not judge us. It's just a bot. Its job is to answer whatever we are asking.''} (P9, Voice Dialogue, Bilingual)
\end{quote}

The tutor's responsive guidance, encouragement, and perceived lack of judgment may have led to higher interest and self-efficacy reported in the dialogic conditions. Students appreciated the opportunities to ask questions, receive immediate feedback, and continue working through uncertainty. However, these affective benefits did not translate into reliably greater knowledge gains.

\subsubheading{Bilingual dialogue supported interest and self-efficacy, while English retained aspirational value}\label{subsubsec:english-aspiration}

The relationship between language and learning outcomes depended on the activity format. In written response, bilingual and English participants had nearly identical outcomes. The adjusted differences between the two language conditions were only 0.03 points for interest ($p = .71$), 0.02 points for self-efficacy ($p = .80$), and 0.22 points for knowledge ($p = .74$).

A different pattern emerged in the dialogic activities, particularly for interest. On the five-point scale, bilingual participants reported 0.46 points greater interest than English participants in text dialogue ($p < .001$) and 0.65 points greater interest in voice dialogue ($p < .001$). The bilingual advantage also extended to self-efficacy in voice dialogue, where bilingual participants scored 0.45 points higher than English participants ($p = .001$). Thus, language condition made little difference when students completed a written response, but it mattered when they interacted with the AI tutor. Bilingual dialogue was associated with greater interest in both text and voice, as well as greater self-efficacy in voice dialogue.


The interview analysis revealed that even though Kannada was the language easiest for the students to use, it was not the language that the students considered most valuable. Kannada supported immediate comprehension and expression, whereas students associated English with academic success and future career opportunities. P5 captured this distinction:
\begin{quote}
    \textit{``I am comfortable with Kannada, but if I want to learn, English is more preferable since it is better for my future.''} (P5, Voice Dialogue, Bilingual)
\end{quote}

Students also valued English because their course materials and assessments were in English, even though the classroom instruction was often bilingual. As P3 explained,
\begin{quote}
    \textit{``Our entire syllabus is in English. So if we practice in English, it is easy for us to go and write the exam well in English.''} (P3, Written Response, English)
\end{quote}

This created a tension in students’ language preferences. Although using Kannada made interaction easier and more engaging, especially in voice, some students still preferred to interact with the tutor in English. They viewed the activity not only as an opportunity to learn the topic but also as a way to practice and improve their English, which they associated with their broader educational and professional aspirations.

\subsubheading{Students used AI for support but risked overreliance}\label{subsubsec:students-ai-usage}

Some students used external AI tools while completing the learning activity. Among submitted responses, 9.3\% in written response and 7.4\% in text dialogue were flagged as containing AI-generated text. The rates were similar across language conditions. In written responses, 8.6\% of English and 10\% of bilingual responses were flagged. In text dialogue condition, the corresponding rates were 6.8\% and 8.1\%. These flags do not capture all forms of AI use or establish that every flagged response involved the same level of AI reliance. Nevertheless, they provide behavioral evidence that some students directly incorporated AI-generated text into their responses.

P6 (Written Response, English), whose response was flagged as containing AI-generated text, explained in the interview that that they knew the answer but lacked confidence in how to express the answer. They therefore used AI to help formulate their response. Other students similarly described turning to AI when they found questions difficult to understand. In these cases, students used AI strategically to address a specific difficulty, such as interpreting a question or articulating an answer. 

However, the interviews also revealed the risk of this support becoming habitual. Students described AI as a routine part of their everyday lives and sometimes turned to it before attempting to think through a question first. P9 reflected on how this habit extended beyond academic tasks:

\begin{quote}
\textit{``AI has become so much a part of daily usage. People are even asking, `Today is Tuesday, what color do I need to wear?' It has become more like an everyday friend. Rather than asking my friends what to do next, I just use AI. Even if my friends ask me something, I tell them, `I don't know, just use ChatGPT. Just use AI, it will tell you.' We are not even sitting for two minutes and thinking about it.''} (P9, Voice Dialogue, Bilingual)
\end{quote}

P8 similarly described using AI to obtain short assignment answers rather than finding and combining information from other sources:

\begin{quote}
\textit{``If ma'am gives a question, we mostly search in ChatGPT for the answer because it gives it shortly. Otherwise, we have to go to the library, take a book, and find out page by page where the answer is. It's too difficult, so we use ChatGPT to write assignment answers.''} (P8, Voice Dialogue, Bilingual)
\end{quote}

These accounts reveal a tension in how AI supported students in EMI settings. AI could reduce language-related barriers and help students understand questions or express their ideas. Yet its convenience could also encourage students to seek ready-made answers before reasoning through a question themselves. The challenge, therefore, is not simply to provide access to AI, but to design interactions that offer support while preserving students’ effort and opportunities to think independently.
\section{Discussion}

Our findings highlight a growing challenge for written response-based learning activities: as AI becomes embedded in students’ academic routines, a completed written response doesn't reliably provide evidence of students’ own thinking and understanding. As shown in \S\ref{subsubsec:students-ai-usage}, students regularly used AI to find information, understand difficult content, and complete assignments. Some of these uses supported their learning, while others involved relying on AI to produce assignment answers with little effort of their own. These observations reflect broader patterns of both productive and unproductive AI use across educational contexts \cite{leeImpactGenerativeAI2025,chirikovGenerativeAIUse2026,ravseljHigherEducationStudents2025,mohammadiGenerativeAIReshaping2026,johnstonStudentPerspectivesUse2024}. They point to the need for learning activities that elicit students’ reasoning throughout the process rather than treating a completed written response as sufficient evidence of engagement and understanding. Our findings suggest that AI-based dialogic activities offer one way to support this shift.


Students who completed the dialogic activities participated more, spent longer on the activity (\S\ref{subsubsec:higher-engagement}), and reported greater interest and self-efficacy (\S\ref{subsubsec:increased-interest}) than those who completed the written response activity, while showing comparable knowledge gains (\S\ref{subsubsec:knowledge-gain}). Students described how immediate feedback, hints, and encouragement helped them continue when they did not know an answer (\S\ref{subsubsec:increased-interest}). These findings align with prior work showing how conversational AI can support reflection and productive struggle \cite{daryantoConversateSupportingReflective2025,tangGenerativeAIDialogic2026,tangDialogicApproachTransform2024,biswasEncouragingProductiveStruggle2026}. At the same time, fewer students completed dialogic activities (\S\ref{subsubsec:dialogic-less-completion}), and participation varied substantially both within and across the text and voice conditions (\S\ref{sec:impact-engagement-completion}). These variations were driven by affordances of the modalities, students' cognitive and language capabilities, and their professional aspirations. Therefore, redesigning written activities into dialogic activities involves more than simply replacing writing with dialogue or selecting one interaction modality over another. 
Based on our findings, we discuss two design implications: designing for learner control and designing for aspirations beyond ability.

\subsection{Designing for learner control}

Clark and Brennan's framework of grounding in communication provides a useful lens for understanding how voice and text shaped participation in dialogic activities \cite{clarkGroundingCommunication1991}. The framework describes communication as a collaborative process in which participants work to establish mutual understanding. Communication media shape this process through constraints such as audibility (communicating through speech), cotemporality (receiving a contribution as soon as it is being produced), reviewability (revisiting earlier contributions), and revisability (editing a contribution before sharing it). These constraints affect the effort required to establish and maintain mutual understanding between participants.

For some students, the audibility and cotemporality of voice reduced the mechanical effort of typing and supported a more natural conversational flow (\S\ref{subsubsec:voice-interaction}). These benefits were particularly visible when students could draw on their fluency in spoken Kannada. Audibility also allowed the AI-generated voice to convey social and emotional cues that were less readily available through text. These cues could make the interaction feel more \textit{human}, as observed in prior work \cite{rabinAIVoicesFollow2026}. Social Agency Theory suggests that social cues in instructional materials, including the use of voice, can encourage learners to respond socially to the system and may support learning \cite{mayerPrinciplesMultimediaLearning2005,craigReconsideringVoiceEffect2017}. Voice modality therefore aligned particularly well with languages that students could already speak fluently.

However the voice modality also introduced several difficulties. Students found it challenging that voice offered limited reviewability as the students couldn't refer back to the earlier parts of the conversation (\S\ref{subsubsec:voice-interaction}). It also offered limited revisability because students had little opportunity to change an answer before it became part of the exchange (\S\ref{subsubsec:language-intricacies}). These demands became greater when students interacted in English. They had to understand the academic content, process spoken English, retain earlier information, and prepare an English response at the same time. The Transient Information Effect explains how information that disappears as the interaction continues can place additional demands on working memory \cite{singhTransientInformationEffect2012}. These demands can be particularly challenging in EMI settings, where students must process both academic content and a non-native language \cite{swellerCognitiveLoadTheory2011,dejongCognitiveLoadTheory2010,rousselLearningSubjectContent2017a}.

Text dialogue had its own advantages and challenges. Students appreciated how its reviewability allowed them to return to earlier messages when they forgot information or lost track of the conversation. Its revisability also gave them time to formulate, inspect, and change an English response before sending it (\S\ref{subsubsec:language-intricacies}). These features translated into how students experienced stress during learning. Prior work with non-native English speakers similarly found that text chat reduced learner anxiety compared with voice chat, although both supported learning \cite{namaziandostSynchronousCMCText2022}. However, typing in vernacular languages also created their own challenges. Students in our study found Kannada more difficult to type than to speak. Prior work has similarly documented the challenges of text entry in Indic languages \cite{reitmaierOpportunitiesChallengesAutomatic2022,balkrishanMakingBreakingUserUsage2016,ghoshTextEntryIndian2014,karusalaOnlyIfYou2018}.

\parabold{Design implication: Give learners control over modality, information, and pace}
Based on these findings, we propose that AI-based dialogic activities should offer both voice and text rather than require students to use one modality throughout the activity. Students should be able to choose the modality that best supports their participation at a given moment and switch modalities as the language and demands of the interaction change. AI responses in voice dialogue could also remain available as text, allowing students to review an explanation without having to remember it. Prior work has shown that providing written speech-to-text content alongside spoken content can improve learning and reduce cognitive load for non-native English speakers, particularly those with lower English proficiency \cite{huangInvestigatingEffectivenessSpeechtotext2016}.

Learner control should also extend to the pace of the conversation. Students should be able to indicate when they have finished responding and are ready for the next response. For example, a design similar to \textit{push-to-talk} could allow a student to pause, think, and formulate an answer without the system treating that delay as the end of the turn. Prior work on push-to-talk communication has shown that it can support interactions in which delays between turns are acceptable \cite{woodruffHowPushtotalkMakes2003}. For learning activities, such a control can protect the time students need for reflection. Further, when a contribution is misunderstood, the interface should also allow students to inspect what the system recognized, correct it, and continue from the point of breakdown. Learner control should therefore include how students respond, what conversational information remains available, and when the conversation moves forward.

Meaningful learner control also depends on reliable speech and language technologies. In our study, the AI tutor sometimes misunderstood a student’s response, causing the conversation to break down (\S\ref{subsubsec:language-intricacies}). Students then had to correct the misunderstanding or repeat themselves, diverting attention from the academic task to repairing the interaction \cite{levineHowHighSchool2023}. Prior work has similarly documented disparities in speech-recognition performance across accents and languages \cite{koeneckeRacialDisparitiesAutomated2020,fengInclusiveAutomaticSpeech2024,nakatumba-nabendeSystematicLiteratureReview2025}, as well as errors in language technologies designed for low-resource languages \cite{bhagat_tales_2026,varuvel_dennison_shiksha_2026,varuvel_dennison_designing_2026}. Because such errors cannot always be prevented, systems must also make them visible and easy to repair. Voice interfaces could display an on-screen transcription of the student’s response, allow the student to correct it before the tutor proceeds, and provide a simple way to switch to text when speech recognition repeatedly fails.

\subsection{Designing for aspirations beyond ability}

Aspiration-based design helps us interpret how students made choices about language in our study. It argues that design should account for what people hope to achieve, along with what they can do in their present circumstances \cite{kumarAspirationsbasedDesign2019,toyamaNeedsAspirationsInformation2018}. In our study, students evaluated a language not only by how easily it allowed them to participate, but also by the educational and professional opportunities they associated with it (\S\ref{subsubsec:english-aspiration}).

The importance of supporting students’ current language capabilities was most apparent in voice dialogue. Compared with English voice dialogue, bilingual voice dialogue had fewer turns showing articulation difficulty, more turns demonstrating understanding, and fewer conversations abandoned before completion (\S\ref{subsubsec:language-intricacies}).
Students in the bilingual conditions reported greater interest than those in the English-only conditions across both text and voice dialogue. In the voice condition, students who interacted bilingually also reported greater self-efficacy than those who interacted only in English; this difference did not emerge in text dialogue (\S\ref{subsubsec:english-aspiration}).
Bilingual interaction was therefore particularly valuable when students had to understand and respond during a real-time voice conversation.

Yet the language that best supported students’ immediate participation was not always the language they preferred. Although Kannada made it easier for students to understand questions and express their ideas, many still preferred to interact in English (\S\ref{subsubsec:english-aspiration}). Students emphasized that English was the language of their curriculum and examinations, and associated it with better career prospects. Prior work in multilingual HCI has similarly shown how the use of English can be connected to aspirations for educational and economic mobility \cite{karusalaOnlyIfYou2018}. Students' use of Kannada and their preference for English were therefore not contradictory: Kannada helped them participate in the present, while English represented capabilities and opportunities they wanted to develop.

Designing only for English could prevent students with lower proficiency from participating fully in learning activities. 
This is particularly important in EMI settings, where students are already challenged to learn academic content through a non-native language in which they may have limited proficiency \cite{treffers-dallerHowReadyAre2022,jhaPublicPolicyDilemma2021,endowInferiorOutcomesLearning2018,bhattacharyaMediatingInequalitiesExploring2013}. Conversely, designing only around the language students currently find easiest could limit opportunities to develop English proficiency. The design challenge is therefore not to determine whether English or native language is universally better, but to use native-language support to make English-medium learning more accessible while preserving meaningful opportunities to practice English. 

\parabold{Design implication: Use translanguaging to connect present abilities with future aspirations}
Translanguaging offers one way to address this challenge. It does not treat a multilingual learner's languages as separate resources that must be used one at a time. Instead, learners can draw on their languages together to understand concepts, formulate ideas, and communicate their understanding \cite{garcia2009bilingual,baileyTranslanguagingConceptualUnderpinnings2025,vogelTranslanguaging}. In our context, Kannada could help a student clarify a question or develop an initial explanation, while English could remain part of how they communicate what they have understood. Kannada would then serve as scaffolding for continued participation rather than as a substitute for English.

Multilingual capabilities of AI can allow this scaffolding to be provided within an academic activity. Recent studies have begun to examine how AI can support translanguaging and help learners move between languages \cite{tziridesMultiliteraciesTranslanguagingDigital2026,kuzuAISupportedTranslanguagingProcesses2026,waluyoTeacherLedApproachesStudentInitiated2025,jeonGenerativeAIIts2025}. Such uses may also support students' academic engagement and emotional experiences \cite{derakhshanLeveragingTranslanguagingGenAIenhanced2026}. 
Much of this work, however, focuses on language learning. Our findings show that translanguaging is also relevant when students are learning academic content through a non-native language. AI-based dialogues could allow students to use their native language to work through curricular concepts while continuing to practice academic English. This approach could connect students' interest in conversational English practice with institutions' goals of supporting curriculum-aligned learning \cite{shashidharaVoiceBasedChatbotsEnglish2026}.
\section{Limitations and Future Work}
While our study was conducted through a real-world deployment, several limitations remain. First, the learning context in which the study was conducted was optional and low-stakes for students, which may have influenced their engagement and completion. We deliberately chose this setting to minimize potential harms to students' academic progress and to allow students to participate without consequences for their grades. However, the optional nature of the activity may have contributed to the observed attrition. Of the 305 students who enrolled, 173 completed the study, with completion rates differing across interaction modalities. Although the modality pattern persisted in the second deployment, where completion rates were substantially higher, differential attrition between experimental conditions may nevertheless have influenced the students represented in our outcome analyses. Future work could examine these interventions in required or graded learning contexts to determine whether the observed patterns persist when students have stronger external incentives to participate, while also investigating the factors that contribute to non-completion. Larger-scale studies would further help provide sufficient statistical power to detect potentially smaller effects and examine whether effects vary across student characteristics and learning contexts.

Second, although the study was conducted in a specific educational context in Karnataka, the deployment setting reflects an important and relatively common form of English-medium education in which students navigate between English and their local language. Nevertheless, the study was conducted within a single state and with a particular linguistic population, limiting the extent to which the findings can be generalized across diverse educational and linguistic contexts. Future work could replicate the study at a larger scale across multiple states, institutions, and linguistic communities to assess the robustness and transferability of the findings. Such studies could also provide larger and more diverse samples for examining how language, interaction modality, and students' educational and linguistic backgrounds jointly shape their experiences with conversational AI.
\section{Conclusion}
This study examined how activity format, interaction modality, and language shape engagement and learning among 305 Kannada-speaking students in English-medium institutions. Compared with written responses, dialogic activities with AI elicited greater participation, time on task, interest, and self-efficacy among students who completed them, while producing comparable knowledge gains and lower overall completion. Within dialogic activities, language shaped participation differently across modalities. Bilingual support was particularly beneficial in voice interaction, where it reduced articulation difficulties, supported demonstrations of understanding, and made students less likely to abandon the conversation. At the same time, students continued to value English because of its role in their education and future opportunities. Further, these findings position dialogic AI not as a universally better replacement for writing, but as a distinct learning format whose benefits depend on whether students can sustain and complete the interaction.

These findings show that responding to AI-driven transformations in education involves more than replacing written activities with dialogue or selecting a single modality or language. Students in EMI contexts navigate the demands of learning academic content, participating through a non-native language, and developing the language capabilities connected to their aspirations. This work can support educators, edtech designers, and AI developers in thinking about how learning technologies might respond to these intersecting needs. In particular, it highlights the importance of designing AI-based dialogic learning experiences that make participation more accessible while preserving students’ opportunities to think, express their own understanding, and develop the capabilities they value. Overall, supporting students through the changing landscape of learning with AI requires technologies that account for both how students are presently able to participate and how they hope to learn and grow.

\bibliographystyle{ACM-Reference-Format}
\bibliography{related-work,methods,references}

@article{hornbergerMultinationalAssessmentAI2025,
  title = {A Multinational Assessment of {{AI}} Literacy among University Students in {{Germany}}, the {{UK}}, and the {{US}}},
  author = {Hornberger, Marie and Bewersdorff, Arne and Schiff, Daniel S. and Nerdel, Claudia},
  year = 2025,
  journal = {Computers in Human Behavior: Artificial Humans},
  volume = {4},
  pages = {100132},
  issn = {2949-8821},
  doi = {10.1016/j.chbah.2025.100132}
}

@misc{zhouTutorMoveTaxonomy2026,
  title = {Tutor {{Move Taxonomy}}: {{A Theory-Aligned Framework}} for {{Analyzing Instructional Moves}} in {{Tutoring}}},
  shorttitle = {Tutor {{Move Taxonomy}}},
  author = {Zhou, Zhuqian and Vanacore, Kirk and Thompson, Tamisha and {St John}, Jennifer and Kizilcec, Ren{\'e}},
  year = 2026,
  eprint = {2603.05778},
  archiveprefix = {arXiv},
  primaryclass = {cs.CY},
  url = {https://arxiv.org/abs/2603.05778}
}

@inproceedings{sureshTalkMovesDataset2022,
  title = {The {{TalkMoves}} Dataset: {{K-12}} Mathematics Lesson Transcripts Annotated for Teacher and Student Discursive Moves},
  shorttitle = {The {{TalkMoves}} Dataset},
  author = {Suresh, Abhijit and Jacobs, Jennifer and Harty, Charis and Perkoff, Margaret and Martin, James H. and Sumner, Tamara},
  year = 2022,
  month = jun,
  booktitle = {Proceedings of the Thirteenth Language Resources and Evaluation Conference},
  publisher = {European Language Resources Association},
  address = {Marseille, France},
  pages = {4654--4662},
  url = {https://aclanthology.org/2022.lrec-1.497/}
}

@article{stolckeDialogueActModeling2000,
  title = {Dialogue {{Act Modeling}} for {{Automatic Tagging}} and {{Recognition}} of {{Conversational Speech}}},
  author = {Stolcke, Andreas and Ries, Klaus and Coccaro, Noah and Shriberg, Elizabeth and Bates, Rebecca and Jurafsky, Daniel and Taylor, Paul and Martin, Rachel and Van {Ess-Dykema}, Carol and Meteer, Marie},
  year = 2000,
  journal = {Computational Linguistics},
  volume = {26},
  number = {3},
  pages = {339--373},
  doi = {10.1162/089120100561737}
}

@inproceedings{varuvel_dennison_designing_2026,
	address = {New York, NY, USA},
	series = {{CHI} '26},
	title = {Designing {Culturally} {Aligned} {AI} {Systems} {For} {Social} {Good} in {Non}-{Western} {Contexts}},
	isbn = {979-8-4007-2278-3},
	url = {https://dl.acm.org/doi/10.1145/3772318.3791513},
	doi = {10.1145/3772318.3791513},
	urldate = {2026-05-06},
	booktitle = {Proceedings of the 2026 {CHI} {Conference} on {Human} {Factors} in {Computing} {Systems}},
	publisher = {Association for Computing Machinery},
	author = {Varuvel Dennison, Deepak and Jain, Mohit and Ganu, Tanuja and Vashistha, Aditya},
	month = apr,
	year = {2026},
	pages = {1--22},
}

@article{varuvel_dennison_shiksha_2026,
	title = {Shiksha {Copilot}: {Teacher}-{AI} {Collaboration} for {Curating} and {Customizing} {Lesson} {Plans} in {Low}-{Resource} {Schools}},
	volume = {10},
	shorttitle = {Shiksha {Copilot}},
	url = {https://dl.acm.org/doi/10.1145/3788074},
	doi = {10.1145/3788074},
	number = {2},
	urldate = {2026-08-24},
	journal = {Proceedings of the ACM on Human-Computer Interaction},
	author = {Varuvel Dennison, Deepak and Ahtisham, Bakhtawar and Chourasia, Kavyansh and Arora, Nirmit and Singh, Rahul and Kizilcec, René F. and Nambi, Akshay and Ganu, Tanuja and Vashistha, Aditya},
	month = may,
	year = {2026},
	pages = {CSCW038:1--CSCW038:47},
}

@inproceedings{bhagat_tales_2026,
	address = {New York, NY, USA},
	series = {{CHI} '26},
	title = {{TALES}: {A} {Taxonomy} and {Analysis} of {Cultural} {Representations} in {LLM}-generated {Stories}},
	isbn = {979-8-4007-2278-3},
	shorttitle = {{TALES}},
	url = {https://dl.acm.org/doi/10.1145/3772318.3790519},
	doi = {10.1145/3772318.3790519},
	urldate = {2026-08-24},
	booktitle = {Proceedings of the 2026 {CHI} {Conference} on {Human} {Factors} in {Computing} {Systems}},
	publisher = {Association for Computing Machinery},
	author = {Bhagat, Kirti and Bhatt, Shaily and Velagapudi, Athul and Vashistha, Aditya and Dave, Shachi and Pruthi, Danish},
	month = apr,
	year = {2026},
	pages = {1--26},
}

@article{feredayDemonstratingRigorUsing2006,
    title = {Demonstrating {Rigor} {Using} {Thematic} {Analysis}: {A} {Hybrid} {Approach} of {Inductive} and {Deductive} {Coding} and {Theme} {Development}},
    volume = {5},
    issn = {1609-4069},
    shorttitle = {Demonstrating {Rigor} {Using} {Thematic} {Analysis}},
    url = {https://doi.org/10.1177/160940690600500107},
    doi = {10.1177/160940690600500107},
    language = {en},
    number = {1},
    urldate = {2024-12-26},
    journal = {International Journal of Qualitative Methods},
    publisher = {SAGE Publications Inc},
    author = {Fereday, Jennifer and Muir-Cochrane, Eimear},
    month = mar,
    year = {2006},
    pages = {80--92},
}

@incollection{mayerPrinciplesMultimediaLearning2005,
  title = {Principles of {{Multimedia Learning Based}} on {{Social Cues}} : {{Personalization}}, {{Voice}}, and {{Image Principles}}},
  shorttitle = {Principles of {{Multimedia Learning Based}} on {{Social Cues}}},
  booktitle = {The {{Cambridge Handbook}} of {{Multimedia Learning}}},
  author = {Mayer, Richard E.},
  editor = {Mayer, Richard},
  date = {2005-08-15},
  edition = {1},
  pages = {201--212},
  publisher = {Cambridge University Press},
  doi = {10.1017/CBO9780511816819.014},
  url = {https://www.cambridge.org/core/product/identifier/CBO9780511816819A021/type/book_part},
  urldate = {2026-08-21},
  isbn = {978-0-521-83873-3 978-0-521-54751-2 978-0-511-81681-9},
  langid = {english}
}

@article{craigReconsideringVoiceEffect2017,
  title = {Reconsidering the Voice Effect When Learning from a Virtual Human},
  author = {Craig, Scotty D. and Schroeder, Noah L.},
  date = {2017-11-01},
  journaltitle = {Computers \& Education},
  shortjournal = {Comput. Educ.},
  volume = {114},
  pages = {193--205},
  issn = {0360-1315},
  doi = {10.1016/j.compedu.2017.07.003},
  url = {https://www.sciencedirect.com/science/article/pii/S0360131517301653},
  urldate = {2026-08-21}
}

@inproceedings{shashidharaVoiceBasedChatbotsEnglish2026,
  title = {Voice-{{Based Chatbots}} for {{English Speaking Practice}} in {{Multilingual Low-Resource Indian Schools}}: {{A Multi-Stakeholder Study}}},
  shorttitle = {Voice-{{Based Chatbots}} for {{English Speaking Practice}} in {{Multilingual Low-Resource Indian Schools}}},
  booktitle = {Proceedings of the 2026 {{CHI Conference}} on {{Human Factors}} in {{Computing Systems}}},
  author = {Shashidhara, Sneha and Chi, Vivienne Bihe and Singh, Abhay P and Ungar, Lyle and Guntuku, Sharath Chandra},
  date = {2026-04-13},
  series = {{{CHI}} '26},
  pages = {1--19},
  publisher = {Association for Computing Machinery},
  location = {New York, NY, USA},
  doi = {10.1145/3772318.3791266},
  url = {https://dl.acm.org/doi/10.1145/3772318.3791266},
  urldate = {2026-08-22},
  isbn = {979-8-4007-2278-3}
}

@article{namaziandostSynchronousCMCText2022,
  title = {Synchronous {{CMC}} Text Chat versus Synchronous {{CMC}} Voice Chat: Impacts on {{EFL}} Learners’ Oral Proficiency and Anxiety},
  shorttitle = {Synchronous {{CMC}} Text Chat versus Synchronous {{CMC}} Voice Chat},
  author = {Namaziandost, Ehsan and Razmi, Mohammad Hasan and Hernández, Ronald M. and Ocaña-Fernández, Yolvi and Khabir, Masoud},
  date = {2022-08-08},
  journaltitle = {Journal of Research on Technology in Education},
  shortjournal = {J. Res. Technol. Educ.},
  volume = {54},
  number = {4},
  pages = {599--616},
  publisher = {Routledge},
  issn = {1539-1523},
  doi = {10.1080/15391523.2021.1906362},
  url = {https://doi.org/10.1080/15391523.2021.1906362},
  urldate = {2026-08-22}
}

@inproceedings{reitmaierOpportunitiesChallengesAutomatic2022,
  title = {Opportunities and {{Challenges}} of {{Automatic Speech Recognition Systems}} for {{Low-Resource Language Speakers}}},
  booktitle = {Proceedings of the 2022 {{CHI Conference}} on {{Human Factors}} in {{Computing Systems}}},
  author = {Reitmaier, Thomas and Wallington, Electra and Kalarikalayil Raju, Dani and Klejch, Ondrej and Pearson, Jennifer and Jones, Matt and Bell, Peter and Robinson, Simon},
  date = {2022-04-29},
  series = {{{CHI}} '22},
  pages = {1--17},
  publisher = {Association for Computing Machinery},
  location = {New York, NY, USA},
  doi = {10.1145/3491102.3517639},
  url = {https://dl.acm.org/doi/10.1145/3491102.3517639},
  urldate = {2026-08-22},
  isbn = {978-1-4503-9157-3}
}

@inproceedings{balkrishanMakingBreakingUserUsage2016,
  title = {Making and {{Breaking}} the {{User-Usage Model}}: {{WhatsApp Adoption Amongst Emergent Users}} in {{India}}},
  shorttitle = {Making and {{Breaking}} the {{User-Usage Model}}},
  booktitle = {Proceedings of the 8th {{Indian Conference}} on {{Human-Computer Interaction}}},
  author = {Balkrishan, Devanuj Kanta and Joshi, Anirudha and Rajendran, Chandni and Nizam, Nazreen and Parab, Chinmay and Devkar, Sujit},
  date = {2016-12-07},
  series = {{{IndiaHCI}} '16},
  pages = {52--63},
  publisher = {Association for Computing Machinery},
  location = {New York, NY, USA},
  doi = {10.1145/3014362.3014367},
  url = {https://dl.acm.org/doi/10.1145/3014362.3014367},
  urldate = {2026-08-22},
  isbn = {978-1-4503-4863-8}
}

@inproceedings{ghoshTextEntryIndian2014,
  title = {Text {{Entry}} in {{Indian Languages}} on {{Mobile}}: {{User Perspectives}}},
  shorttitle = {Text {{Entry}} in {{Indian Languages}} on {{Mobile}}},
  booktitle = {Proceedings of the 6th {{Indian Conference}} on {{Human-Computer Interaction}}},
  author = {Ghosh, Sanjay and Joshi, Anirudha},
  date = {2014-12-07},
  series = {{{IndiaHCI}} '14},
  pages = {55--63},
  publisher = {Association for Computing Machinery},
  location = {New York, NY, USA},
  doi = {10.1145/2676702.2676710},
  url = {https://dl.acm.org/doi/10.1145/2676702.2676710},
  urldate = {2026-08-22},
  isbn = {978-1-4503-3218-7}
}

@article{huangInvestigatingEffectivenessSpeechtotext2016,
  title = {Investigating the Effectiveness of Speech-to-Text Recognition Applications on Learning Performance and Cognitive Load},
  author = {Huang, Yueh-Min and Shadiev, Rustam and Hwang, Wu-Yuin},
  date = {2016-10-01},
  journaltitle = {Computers \& Education},
  shortjournal = {Comput. Educ.},
  volume = {101},
  pages = {15--28},
  issn = {0360-1315},
  doi = {10.1016/j.compedu.2016.05.011},
  url = {https://www.sciencedirect.com/science/article/pii/S036013151630121X},
  urldate = {2026-08-22}
}

@article{levineHowHighSchool2023,
  title = {How High School Students Used Speech-to-Text as a Composition Tool},
  author = {Levine, Sarah and Hsieh, Hsiaolin and Southerton, Emily and Silverman, Rebecca},
  date = {2023-06-01},
  journaltitle = {Computers and Composition},
  shortjournal = {Comput. Compos.},
  volume = {68},
  pages = {102775},
  issn = {8755-4615},
  doi = {10.1016/j.compcom.2023.102775},
  url = {https://www.sciencedirect.com/science/article/pii/S8755461523000269},
  urldate = {2026-08-22}
}

@article{singhTransientInformationEffect2012,
  title = {The {{Transient Information Effect}}: {{Investigating}} the {{Impact}} of {{Segmentation}} on {{Spoken}} and {{Written}} Text},
  shorttitle = {The {{Transient Information Effect}}},
  author = {Singh, Anne-Marie and Marcus, Nadine and Ayres, Paul},
  date = {2012},
  journaltitle = {Applied Cognitive Psychology},
  shortjournal = {Appl. Cogn. Psychol.},
  volume = {26},
  number = {6},
  pages = {848--853},
  issn = {1099-0720},
  doi = {10.1002/acp.2885},
  url = {https://onlinelibrary.wiley.com/doi/abs/10.1002/acp.2885},
  urldate = {2026-08-22},
  langid = {english}
}

@inproceedings{woodruffHowPushtotalkMakes2003,
  title = {How Push-to-Talk Makes Talk Less Pushy},
  booktitle = {Proceedings of the 2003 {{ACM International Conference}} on {{Supporting Group Work}}},
  author = {Woodruff, Allison and Aoki, Paul M.},
  date = {2003-11-09},
  series = {{{GROUP}} '03},
  pages = {170--179},
  publisher = {Association for Computing Machinery},
  location = {New York, NY, USA},
  doi = {10.1145/958160.958187},
  url = {https://dl.acm.org/doi/10.1145/958160.958187},
  urldate = {2026-08-22},
  isbn = {978-1-58113-693-7}
}

@article{rabinAIVoicesFollow2026,
  title = {Do {{AI}} Voices Follow Social Nuances? {{The}} Case of Politeness and Speech Rate},
  shorttitle = {Do {{AI}} Voices Follow Social Nuances?},
  author = {Rabin, Eyal and Elyoseph, Zohar and Israel-Fishelson, Rotem and Dali, Adi and Nussinson, Ravit},
  date = {2026-03-01},
  journaltitle = {Computers in Human Behavior: Artificial Humans},
  shortjournal = {Comput. Hum. Behav. Artif. Hum.},
  volume = {7},
  pages = {100256},
  issn = {2949-8821},
  doi = {10.1016/j.chbah.2026.100256},
  url = {https://www.sciencedirect.com/science/article/pii/S2949882126000071},
  urldate = {2026-08-23}
}

@article{koeneckeRacialDisparitiesAutomated2020,
  title = {Racial Disparities in Automated Speech Recognition},
  author = {Koenecke, Allison and Nam, Andrew and Lake, Emily and Nudell, Joe and Quartey, Minnie and Mengesha, Zion and Toups, Connor and Rickford, John R. and Jurafsky, Dan and Goel, Sharad},
  date = {2020-04-07},
  journaltitle = {Proceedings of the National Academy of Sciences},
  shortjournal = {Proc. Natl. Acad. Sci.},
  volume = {117},
  number = {14},
  pages = {7684--7689},
  publisher = {Proceedings of the National Academy of Sciences},
  doi = {10.1073/pnas.1915768117},
  url = {https://www.pnas.org/doi/full/10.1073/pnas.1915768117},
  urldate = {2026-08-24}
}

@article{fengInclusiveAutomaticSpeech2024,
  title = {Towards Inclusive Automatic Speech Recognition},
  author = {Feng, Siyuan and Halpern, Bence Mark and Kudina, Olya and Scharenborg, Odette},
  date = {2024-03-01},
  journaltitle = {Computer Speech \& Language},
  shortjournal = {Comput. Speech Lang.},
  volume = {84},
  pages = {101567},
  issn = {0885-2308},
  doi = {10.1016/j.csl.2023.101567},
  url = {https://www.sciencedirect.com/science/article/pii/S0885230823000864},
  urldate = {2026-08-24}
}

@article{nakatumba-nabendeSystematicLiteratureReview2025,
  title = {A {{Systematic Literature Review}} on {{Bias Evaluation}} and {{Mitigation}} in {{Automatic Speech Recognition Models}} for {{Low-Resource African Languages}}},
  author = {Nakatumba-Nabende, Joyce and Kagumire, Sulaiman and Kantono, Caroline and Nabende, Peter},
  date = {2025-10-27},
  journaltitle = {ACM Computing Surveys},
  shortjournal = {ACM Comput. Surv.},
  volume = {58},
  number = {4},
  pages = {105:1--105:24},
  issn = {0360-0300},
  doi = {10.1145/3769089},
  url = {https://dl.acm.org/doi/10.1145/3769089},
  urldate = {2026-08-24}
}

@inproceedings{kumarAspirationsbasedDesign2019,
  title = {Aspirations-Based Design},
  booktitle = {Proceedings of the {{Tenth International Conference}} on {{Information}} and {{Communication Technologies}} and {{Development}}},
  author = {Kumar, Neha and Wong-Villacres, Marisol and Karusala, Naveena and Vishwanath, Aditya and Kumar, Arkadeep and Ismail, Azra},
  date = {2019-01-04},
  series = {{{ICTD}} '19},
  pages = {1--11},
  publisher = {Association for Computing Machinery},
  location = {New York, NY, USA},
  doi = {10.1145/3287098.3287117},
  url = {https://dl.acm.org/doi/10.1145/3287098.3287117},
  urldate = {2026-08-23},
  isbn = {978-1-4503-6122-4}
}

@article{toyamaNeedsAspirationsInformation2018,
  title = {From Needs to Aspirations in Information Technology for Development},
  author = {Toyama, Kentaro},
  date = {2018-01-02},
  journaltitle = {Information Technology for Development},
  shortjournal = {Inf. Technol. Dev.},
  volume = {24},
  number = {1},
  pages = {15--36},
  issn = {0268-1102, 1554-0170},
  doi = {10.1080/02681102.2017.1310713},
  url = {https://www.tandfonline.com/doi/full/10.1080/02681102.2017.1310713},
  urldate = {2026-08-24},
  langid = {english}
}

@inproceedings{karusalaOnlyIfYou2018,
  title = {"{{Only}} If You Use {{English}} You Will Get to More Things": {{Using Smartphones}} to {{Navigate Multilingualism}}},
  shorttitle = {"{{Only}} If You Use {{English}} You Will Get to More Things"},
  booktitle = {Proceedings of the 2018 {{CHI Conference}} on {{Human Factors}} in {{Computing Systems}}},
  author = {Karusala, Naveena and Vishwanath, Aditya and Vashistha, Aditya and Kumar, Sunita and Kumar, Neha},
  date = {2018-04-21},
  series = {{{CHI}} '18},
  pages = {1--14},
  publisher = {Association for Computing Machinery},
  location = {New York, NY, USA},
  doi = {10.1145/3173574.3174147},
  url = {https://dl.acm.org/doi/10.1145/3173574.3174147},
  urldate = {2026-08-24},
  isbn = {978-1-4503-5620-6}
}

@article{jhaPublicPolicyDilemma2021,
  title = {Public {{Policy Dilemma}} of {{Choosing}} the {{Medium}} of {{Instruction}} for {{School Education}}: {{A Case}} for {{Questioning Fallacies}} and {{Connecting Objectives}}},
  shorttitle = {Public {{Policy Dilemma}} of {{Choosing}} the {{Medium}} of {{Instruction}} for {{School Education}}},
  author = {Jha, Jyotsna},
  date = {2021-08-23},
  journaltitle = {Journal of Asian Development Research},
  shortjournal = {J. Asian Dev. Res.},
  pages = {2633190X211034697},
  publisher = {SAGE Publications},
  issn = {2633-190X},
  doi = {10.1177/2633190X211034697},
  url = {https://doi.org/10.1177/2633190X211034697},
  urldate = {2026-08-24},
  langid = {english}
}

@article{endowInferiorOutcomesLearning2018,
  title = {Inferior {{Outcomes}}: {{Learning}} in {{Low-cost English-medium Private Schools}}—{{A Survey}} in {{Delhi}} and {{National Capital Region}}},
  shorttitle = {Inferior {{Outcomes}}},
  author = {Endow, Tanuka},
  date = {2018-04-01},
  journaltitle = {Indian Journal of Human Development},
  shortjournal = {Indian J. Hum. Dev.},
  volume = {12},
  number = {1},
  pages = {1--19},
  publisher = {SAGE Publications India},
  issn = {0973-7030},
  doi = {10.1177/0973703018779725},
  url = {https://doi.org/10.1177/0973703018779725},
  urldate = {2026-08-24}
}

@article{baileyTranslanguagingConceptualUnderpinnings2025,
  title = {Translanguaging: {{Conceptual}} Underpinnings of Equity-Oriented Instructional and Assessment Practices with Adolescent Multilingual Learners},
  shorttitle = {Translanguaging},
  author = {Bailey, Alison L. and Abraham, Megan E. and Dong, Juno Yingzhi and Torres, Ines M. and Wang, Yifei Phoebe and Zamora, Edwin and Zhang, Xizi},
  date = {2025-06-01},
  journaltitle = {Linguistics and Education},
  shortjournal = {Linguist. Educ.},
  volume = {87},
  pages = {101418},
  issn = {0898-5898},
  doi = {10.1016/j.linged.2025.101418},
  url = {https://www.sciencedirect.com/science/article/pii/S0898589825000361},
  urldate = {2026-08-24}
}

@incollection{vogelTranslanguaging,
  title = {Translanguaging},
  booktitle = {Oxford {{Research Encyclopedia}} of {{Education}}},
  author = {Vogel, Sara and García, Ofelia},
  editor = {Hytten, Kathy},
  pages = {0},
  publisher = {Oxford University Press},
  doi = {10.1093/acrefore/9780190264093.013.181},
  url = {https://doi.org/10.1093/acrefore/9780190264093.013.181},
  urldate = {2026-08-24},
  isbn = {978-0-19-785142-5}
}

@book{garcia2009bilingual,
  title={Bilingual education in the 21st century: A global perspective},
  author={Garc{\'\i}a, Ofelia and Beardsmore, Hugo Baetens},
  volume={268},
  year={2009},
  publisher={Wiley-Blackwell Malden, MA}
}

@article{derakhshanLeveragingTranslanguagingGenAIenhanced2026,
  title = {Leveraging Translanguaging in {{GenAI-enhanced}} Language Classes: Capturing Its Impact on Multilingual {{English}} Learners’ Achievement Emotions and Academic Engagement through Latent Growth Curve Modelling ({{LGCM}})},
  shorttitle = {Leveraging Translanguaging in {{GenAI-enhanced}} Language Classes},
  author = {Derakhshan, Ali and Wei, Li},
  date = {2026-01-07},
  journaltitle = {International Journal of Multilingualism},
  shortjournal = {Int. J. Multiling.},
  volume = {0},
  number = {0},
  pages = {1--17},
  publisher = {Routledge},
  issn = {1479-0718},
  doi = {10.1080/14790718.2025.2610253},
  url = {https://doi.org/10.1080/14790718.2025.2610253},
  urldate = {2026-08-24}
}

@article{tziridesMultiliteraciesTranslanguagingDigital2026,
  title = {Multiliteracies, Translanguaging and Digital Literacy in Higher Education: Exploring the Role of Generative Artificial Intelligence},
  shorttitle = {Multiliteracies, Translanguaging and Digital Literacy in Higher Education},
  author = {Tzirides, Anastasia Olga},
  date = {2026-04-03},
  journaltitle = {Pedagogies: An International Journal},
  shortjournal = {Pedagog. Int. J.},
  volume = {21},
  number = {2},
  pages = {353--374},
  publisher = {Routledge},
  issn = {1554-480X},
  doi = {10.1080/1554480X.2026.2638241},
  url = {https://doi.org/10.1080/1554480X.2026.2638241},
  urldate = {2026-08-24}
}

@article{kuzuAISupportedTranslanguagingProcesses2026,
  title = {{{AI-Supported Translanguaging Processes}} in {{Primary School}}: {{Empirical Insights}} into {{ChatGPT}}’s {{Role}} in {{Multilingual Interactions}}},
  shorttitle = {{{AI-Supported Translanguaging Processes}} in {{Primary School}}},
  author = {Kuzu, Taha Ertuğrul},
  date = {2026-05-16},
  journaltitle = {Technology, Knowledge and Learning},
  shortjournal = {Technol. Knowl. Learn.},
  issn = {2211-1670},
  doi = {10.1007/s10758-026-09974-7},
  url = {https://doi.org/10.1007/s10758-026-09974-7},
  urldate = {2026-08-24},
  langid = {english}
}

@article{waluyoTeacherLedApproachesStudentInitiated2025,
  title = {Beyond {{Teacher-Led Approaches}}: {{Student-Initiated Translanguaging With Artificial Intelligence Tools}} in {{Foreign Language Acquisition}}},
  shorttitle = {Beyond {{Teacher-Led Approaches}}},
  author = {Waluyo, Budi and Rouaghe, Fouzia},
  date = {2025-07-01},
  journaltitle = {Sage Open},
  volume = {15},
  number = {3},
  pages = {21582440251362998},
  publisher = {SAGE Publications},
  issn = {2158-2440},
  doi = {10.1177/21582440251362998},
  url = {https://doi.org/10.1177/21582440251362998},
  urldate = {2026-08-24},
  langid = {english}
}

@article{jeonGenerativeAIIts2025,
  title = {Generative {{AI}} and Its Dilemmas: Exploring {{AI}} from a Translanguaging Perspective},
  shorttitle = {Generative {{AI}} and Its Dilemmas},
  author = {Jeon, Jaeho and Wei, Li and Tai, Kevin W H and Lee, Seongyong},
  date = {2025-08-01},
  journaltitle = {Applied Linguistics},
  shortjournal = {Appl. Linguist.},
  volume = {46},
  number = {4},
  pages = {709--717},
  issn = {0142-6001},
  doi = {10.1093/applin/amaf049},
  url = {https://doi.org/10.1093/applin/amaf049},
  urldate = {2026-08-24}
}

@inproceedings{leeImpactGenerativeAI2025,
  title = {The {{Impact}} of {{Generative AI}} on {{Critical Thinking}}: {{Self-Reported Reductions}} in {{Cognitive Effort}} and {{Confidence Effects From}} a {{Survey}} of {{Knowledge Workers}}},
  shorttitle = {The {{Impact}} of {{Generative AI}} on {{Critical Thinking}}},
  booktitle = {Proceedings of the 2025 {{CHI Conference}} on {{Human Factors}} in {{Computing Systems}}},
  author = {Lee, Hao-Ping (Hank) and Sarkar, Advait and Tankelevitch, Lev and Drosos, Ian and Rintel, Sean and Banks, Richard and Wilson, Nicholas},
  date = {2025-04-25},
  series = {{{CHI}} '25},
  pages = {1--22},
  publisher = {Association for Computing Machinery},
  location = {New York, NY, USA},
  doi = {10.1145/3706598.3713778},
  url = {https://dl.acm.org/doi/10.1145/3706598.3713778},
  urldate = {2026-08-25},
  isbn = {979-8-4007-1394-1}
}

@article{ravseljHigherEducationStudents2025,
  author = {Ravšelj, Dejan and Keržič, Damijana and Tomaževič, Nina and Umek, Lan and Brezovar, Nejc and Iahad, Noorminshah A. and Abdulla, Ali Abdulla and Akopyan, Anait and Segura, Magdalena Waleska Aldana and AlHumaid, Jehan and Allam, Mohamed Farouk and Alló, Maria and Andoh, Raphael Papa Kweku and Andronic, Octavian and Arthur, Yarhands Dissou and Aydın, Fatih and Badran, Amira and Balbontín-Alvarado, Roxana and Saad, Helmi Ben and Bencsik, Andrea and Benning, Isaac and Besimi, Adrian and Bezerra, Denilson da Silva and Buizza, Chiara and Burro, Roberto and Bwalya, Anthony and Cachero, Cristina and Castillo-Briceno, Patricia and Castro, Harold and Chai, Ching Sing and Charalambous, Constadina and Chiu, Thomas K. F. and Clipa, Otilia and Colombari, Ruggero and Escobedo, Luis José H. Corral and Costa, Elísio and Crețulescu, Radu George and Crispino, Marta and Cucari, Nicola and Dalton, Fergus and Kaya, Meva Demir and Dumić-Čule, Ivo and Dwidienawati, Diena and Ebardo, Ryan and Egbenya, Daniel Lawer and Faris, MoezAlIslam Ezzat and Fečko, Miroslav and Ferrinho, Paulo and Florea, Adrian and Fong, Chun Yuen and Francis, Zoë and Ghilardi, Alberto and González-Fernández, Belinka and Hau, Daniela and Hossain, Md Shamim and Hug, Theo and Inasius, Fany and Ismail, Maryam Jaffar and Jahić, Hatidža and Jessa, Morrison Omokiniovo and Kapanadze, Marika and Kar, Sujita Kumar and Kateeb, Elham Talib and Kaya, Feridun and Khadri, Hanaa Ouda and Kikuchi, Masao and Kobets, Vitaliy Mykolayovych and Kostova, Katerina Metodieva and Krasmane, Evita and Lau, Jesus and Law, Wai Him Crystal and Lazăr, Florin and Lazović-Pita, Lejla and Lee, Vivian Wing Yan and Li, Jingtai and López-Aguilar, Diego Vinicio and Luca, Adrian and Luciano, Ruth Garcia and Machin-Mastromatteo, Juan D. and Madi, Marwa and Manguele, Alexandre Lourenço and Manrique, Rubén Francisco and Mapulanga, Thumah and Marimon, Frederic and Marinova, Galia Ilieva and Mas-Machuca, Marta and Mejía-Rodríguez, Oliva and Meletiou-Mavrotheris, Maria and Méndez-Prado, Silvia Mariela and Meza-Cano, José Manuel and Mirķe, Evija and Mishra, Alpana and Mital, Ondrej and Mollica, Cristina and Morariu, Daniel Ionel and Mospan, Natalia and Mukuka, Angel and Jiménez, Silvana Guadalupe Navarro and Nikaj, Irena and Nisheva, Maria Mihaylova and Nisiforou, Efi and Njiku, Joseph and Nomnian, Singhanat and Nuredini-Mehmedi, Lulzime and Nyamekye, Ernest and Obadić, Alka and Okela, Abdelmohsen Hamed and Olenik-Shemesh, Dorit and Ostoj, Izabela and Peralta-Rizzo, Kevin Javier and Peštek, Almir and Pilav-Velić, Amila and Pires, Dilma Rosanda Miranda and Rabin, Eyal and Raccanello, Daniela and Ramie, Agustine and ur Rashid, Md Mamun and Reuter, Robert A. P. and Reyes, Valentina and Rodrigues, Ana Sofia and Rodway, Paul and Ručinská, Silvia and Sadzaglishvili, Shorena and Salem, Ashraf Atta M. S. and Savić, Gordana and Schepman, Astrid and Shahpo, Samia Mokhtar and Snouber, Abdelmajid and Soler, Emma and Sonyel, Bengi and Stefanova, Eliza and Stone, Anna and Strzelecki, Artur and Tanaka, Tetsuji and Cortes, Carolina Tapia and Teira-Fachado, Andrea and Tilga, Henri and Titko, Jelena and Tolmach, Maryna and Turmudi, Dedi and Varela-Candamio, Laura and Vekiri, Ioanna and Vicentini, Giada and Woyo, Erisher and Yorulmaz, Özlem and Yunus, Said A. S. and Zamfir, Ana-Maria and Zhou, Munyaradzi and Aristovnik, Aleksander},
  title = {Higher Education Students' Perceptions of {ChatGPT}: A Global Study of Early Reactions},
  journal = {PLOS ONE},
  year = {2025},
  volume = {20},
  number = {2},
  pages = {e0315011},
  doi = {10.1371/journal.pone.0315011},
  url = {https://doi.org/10.1371/journal.pone.0315011}
}

@article{mohammadiGenerativeAIReshaping2026,
  title = {Is Generative {{AI}} Reshaping Academic Practices Worldwide? {{A}} Survey of Adoption, Benefits, and Concerns},
  shorttitle = {Is Generative {{AI}} Reshaping Academic Practices Worldwide?},
  author = {Mohammadi, Ehsan and Thelwall, Mike and Cai, Yizhou and Collier, Taylor and Tahamtan, Iman and Eftekhar, Azar},
  date = {2026-01-01},
  journaltitle = {Information Processing \& Management},
  shortjournal = {Inf. Process. Manag.},
  volume = {63},
  number = {1},
  pages = {104350},
  issn = {0306-4573},
  doi = {10.1016/j.ipm.2025.104350},
  url = {https://www.sciencedirect.com/science/article/pii/S0306457325002912},
  urldate = {2026-08-26}
}

@article{johnstonStudentPerspectivesUse2024,
  title = {Student Perspectives on the Use of Generative Artificial Intelligence Technologies in Higher Education},
  author = {Johnston, Heather and Wells, Rebecca F. and Shanks, Elizabeth M. and Boey, Timothy and Parsons, Bryony N.},
  date = {2024-02-08},
  journaltitle = {International Journal for Educational Integrity},
  shortjournal = {Int. J. Educ. Integr.},
  volume = {20},
  number = {1},
  pages = {2},
  issn = {1833-2595},
  doi = {10.1007/s40979-024-00149-4},
  url = {https://doi.org/10.1007/s40979-024-00149-4},
  urldate = {2026-08-26},
  langid = {english}
}

@article{biswasEncouragingProductiveStruggle2026,
  title = {Encouraging Productive Struggle in the Age of Generative {{AI}}},
  author = {Biswas, Andrei and Venkatesaramani, Rajagopal},
  date = {2026-05-08},
  journaltitle = {Patterns},
  volume = {7},
  number = {5},
  pages = {101540},
  issn = {2666-3899},
  doi = {10.1016/j.patter.2026.101540},
  url = {https://www.sciencedirect.com/science/article/pii/S2666389926000498},
  urldate = {2026-08-26}
}

@article{tangDialogicApproachTransform2024,
  title = {A Dialogic Approach to Transform Teaching, Learning \& Assessment with Generative {{AI}} in Secondary Education: A Proof of Concept},
  shorttitle = {A Dialogic Approach to Transform Teaching, Learning \& Assessment with Generative {{AI}} in Secondary Education},
  author = {Tang, Kok-Sing and Cooper, Grant and Rappa, Natasha and Cooper, Martin and Sims, Craig and Nonis, Karen},
  date = {2024-07-02},
  journaltitle = {Pedagogies: An International Journal},
  shortjournal = {Pedagog. Int. J.},
  volume = {19},
  number = {3},
  pages = {493--503},
  publisher = {Routledge},
  issn = {1554-480X},
  doi = {10.1080/1554480X.2024.2379774},
  url = {https://doi.org/10.1080/1554480X.2024.2379774},
  urldate = {2026-08-26}
}

@inproceedings{itoUseAIpoweredRewriting2023,
    address = {New York, NY, USA},
    series = {{UIST} '23},
    title = {Use of an {AI}-powered {Rewriting} {Support} {Software} in {Context} with {Other} {Tools}: {A} {Study} of {Non}-{Native} {English} {Speakers}},
    isbn = {979-8-4007-0132-0},
    shorttitle = {Use of an {AI}-powered {Rewriting} {Support} {Software} in {Context} with {Other} {Tools}},
    url = {https://dl.acm.org/doi/10.1145/3586183.3606810},
    doi = {10.1145/3586183.3606810},
    urldate = {2026-08-28},
    booktitle = {Proceedings of the 36th {Annual} {ACM} {Symposium} on {User} {Interface} {Software} and {Technology}},
    publisher = {Association for Computing Machinery},
    author = {Ito, Takumi and Yamashita, Naomi and Kuribayashi, Tatsuki and Hidaka, Masatoshi and Suzuki, Jun and Gao, Ge and Jamieson, Jack and Inui, Kentaro},
    month = oct,
    year = {2023},
    pages = {1--13},
}

@inproceedings{pratherBreakingProgrammingLanguage2025,
    address = {New York, NY, USA},
    series = {{ACE} '25},
    title = {Breaking the {Programming} {Language} {Barrier}: {Multilingual} {Prompting} to {Empower} {Non}-{Native} {English} {Learners}},
    isbn = {979-8-4007-1425-2},
    shorttitle = {Breaking the {Programming} {Language} {Barrier}},
    url = {https://dl.acm.org/doi/10.1145/3716640.3716649},
    doi = {10.1145/3716640.3716649},
    urldate = {2026-08-28},
    booktitle = {Proceedings of the 27th {Australasian} {Computing} {Education} {Conference}},
    publisher = {Association for Computing Machinery},
    author = {Prather, James and Reeves, Brent N and Denny, Paul and Leinonen, Juho and MacNeil, Stephen and Luxton-Reilly, Andrew and Orvalho, João and Alipour, Amin and Alfageeh, Ali and Amarouche, Thezyrie and Kimmel, Bailey and Wright, Jared and Blake, Musa and Barbre, Gweneth},
    month = apr,
    year = {2025},
    pages = {74--84},
}

@inproceedings{kavuriFurtherStudyLeveraging2026,
    address = {New York, NY, USA},
    series = {{ACE} '26},
    title = {Further {Study} of {Leveraging} {LLM} {Tutoring} {Systems} for {Non}-{Native} {English} {Speakers} in {Introductory} {Computer} {Science}},
    isbn = {979-8-4007-2352-0},
    url = {https://dl.acm.org/doi/10.1145/3786228.3786250},
    doi = {10.1145/3786228.3786250},
    urldate = {2026-05-11},
    booktitle = {Proceedings of the 28th {Australasian} {Computing} {Education} {Conference}},
    publisher = {Association for Computing Machinery},
    author = {Kavuri, Sriharsha and Delgado-Diaz, Danae and Ortiz, Kristhian and Chao, Alex and Villegas Molina, Ismael and Ochoa, Benjamin and Porter, Leo},
    month = mar,
    year = {2026},
    pages = {148--157},
}

@article{tienExploringAImediatedLinguistic2025,
    title = {Exploring {AI}-mediated linguistic and cognitive support for non-native {English} speakers in {English}-only higher education},
    volume = {8},
    copyright = {Copyright (c) 2025 Ching-Yi Tien},
    issn = {2621-6485},
    url = {http://journal2.uad.ac.id/index.php/eltej/article/view/14228},
    doi = {10.12928/eltej.v8i3.14228},
    language = {en},
    number = {3},
    urldate = {2026-08-29},
    journal = {English Language Teaching Educational Journal},
    author = {Tien, Ching-Yi and Haji-Othman, Noor Azam},
    month = dec,
    year = {2025},
    pages = {164--176},
}

@article{taylorTranslanguagingAIpoweredTools2026,
    title = {Translanguaging with {AI}-powered tools and {English} for specific academic purposes: {Chinese} international students’ experiences in {Anglophone} {English} medium instruction higher education},
    volume = {15},
    copyright = {De Gruyter expressly reserves the right to use all content for commercial text and data mining within the meaning of Section 44b of the German Copyright Act.},
    issn = {2191-933X},
    shorttitle = {Translanguaging with {AI}-powered tools and {English} for specific academic purposes},
    url = {https://www.degruyterbrill.com/document/doi/10.1515/jelf-2025-0025/html?lang=en},
    doi = {10.1515/jelf-2025-0025},
    language = {en},
    number = {1},
    urldate = {2026-08-29},
    journal = {Journal of English as a Lingua Franca},
    publisher = {De Gruyter Mouton},
    author = {Taylor, Pimsiri and Gunn, Andrew},
    month = apr,
    year = {2026},
    pages = {1--26},
}

@article{zhouSupportingEnglishSpeaking2025,
    title = {Supporting english speaking practice in higher education: the impact of {AI} chatbot-integrated mobile-assisted blended learning framework},
    volume = {30},
    issn = {1573-7608},
    shorttitle = {Supporting english speaking practice in higher education},
    url = {https://doi.org/10.1007/s10639-025-13401-2},
    doi = {10.1007/s10639-025-13401-2},
    language = {en},
    number = {10},
    urldate = {2026-08-29},
    journal = {Education and Information Technologies},
    author = {Zhou, Qing and Hashim, Harwati and Sulaiman, Nur Ainil},
    month = jul,
    year = {2025},
    pages = {14629--14660},
}

@inproceedings{chroquiAIPoweredEMIClassroom2025,
    address = {Cham},
    title = {{AI}-{Powered} {EMI} {Classroom}: {Enhancing} {Students}’ {Academic} {Understanding} and {Engagement}},
    isbn = {978-3-031-98476-1},
    shorttitle = {{AI}-{Powered} {EMI} {Classroom}},
    doi = {10.1007/978-3-031-98476-1_29},
    language = {en},
    booktitle = {The {Second} {International} {Symposium} on {Generative} {AI} and {Education} ({ISGAIE}’2025)},
    publisher = {Springer Nature Switzerland},
    author = {Chroqui, Razane},
    editor = {González Vallejo, Rubén and Moukhliss, Ghizlane and Schaeffer, Elisa and Paliktzoglou, Vasileios},
    year = {2025},
    pages = {366--379},
}

@article{connellpenskyGenerativeAIsImpact2025,
    title = {Generative {AI}’s {Impact} on {Graduate} {Student} {Professional} {Writing} {Productivity} and {Quality}},
    volume = {35},
    issn = {1560-4306},
    url = {https://doi.org/10.1007/s40593-025-00528-z},
    doi = {10.1007/s40593-025-00528-z},
    language = {en},
    number = {6},
    urldate = {2026-08-29},
    journal = {International Journal of Artificial Intelligence in Education},
    author = {Connell Pensky, Allison E. and Usdan, Jordan H. and Chang, Harley},
    month = dec,
    year = {2025},
    pages = {4057--4082},
}

@inproceedings{halaniExploringEffectivenessMultilingual2025,
    address = {Cham},
    title = {Exploring the {Effectiveness} of a {Multilingual} {Generative} {AI} {Programming} {Chatbot} for {Vernacular} {Medium} {CS} {Students}},
    isbn = {978-3-031-84391-4},
    doi = {10.1007/978-3-031-84391-4_4},
    language = {en},
    booktitle = {Computing {Education} {Research}},
    publisher = {Springer Nature Switzerland},
    author = {Halani, Avval and Prasad, Prajish and Sane, Aamod},
    editor = {Karkare, Amey and Prasad, Prajish and Raman, Arun},
    year = {2025},
    pages = {44--57},
}

@article{wambsganssDesigningConversationalEvaluation2022,
    title = {Designing {Conversational} {Evaluation} {Tools}: {A} {Comparison} of {Text} and {Voice} {Modalities} to {Improve} {Response} {Quality} in {Course} {Evaluations}},
    volume = {6},
    shorttitle = {Designing {Conversational} {Evaluation} {Tools}},
    url = {https://dl.acm.org/doi/10.1145/3555619},
    doi = {10.1145/3555619},
    number = {CSCW2},
    urldate = {2026-08-29},
    journal = {Proceedings of the ACM on Human-Computer Interaction},
    author = {Wambsganss, Thiemo and Zierau, Naim and Söllner, Matthias and Käser, Tanja and Koedinger, Kenneth R. and Leimeister, Jan Marco},
    month = nov,
    year = {2022},
    pages = {506:1--506:27},
}

@article{schlimbachEMBARKINGINTERACTIVELEARNING,
    title = {{EMBARKING} {ON} {AN} {INTERACTIVE} {LEARNING} {JOURNEY}: {EXPLORING} {THE} {INTERACTION} {VALUE} {OF} {VOICEBOTS} {VERSUS} {CHATBOTS}},
    language = {en},
    author = {Schlimbach, Ricarda and Richter, Janine and Grogorick, Linda},
}

@article{wangImpactDifferentConversational2024,
    title = {The impact of different conversational generative {AI} chatbots on {EFL} learners: {An} analysis of willingness to communicate, foreign language speaking anxiety, and self-perceived communicative competence},
    volume = {127},
    issn = {0346-251X},
    shorttitle = {The impact of different conversational generative {AI} chatbots on {EFL} learners},
    url = {https://www.sciencedirect.com/science/article/pii/S0346251X24003154},
    doi = {10.1016/j.system.2024.103533},
    urldate = {2026-08-29},
    journal = {System},
    author = {Wang, Chenghao and Zou, Bin and Du, Yiran and Wang, Zixun},
    month = dec,
    year = {2024},
    pages = {103533},
}

@article{abdelradySpeakingAIImpact2026,
    title = {Speaking with {AI}: the impact of microsoft 365 copilot voice chat on {EFL} learners’ fluency, interactional competence, and autonomy},
    volume = {11},
    issn = {2504-284X},
    shorttitle = {Speaking with {AI}},
    url = {https://www.frontiersin.org/journals/education/articles/10.3389/feduc.2026.1851080/full},
    doi = {10.3389/feduc.2026.1851080},
    language = {English},
    urldate = {2026-08-29},
    journal = {Frontiers in Education},
    publisher = {Frontiers},
    author = {Abdelrady, Abbas Hussein and Khalil, Abdelhamid A. and Ali El Deen, Abdul Aziz Mohamed Mohamed and Alnofal, Abdulaziz Ibrahim S. and Mohamed, Amr M.},
    month = jul,
    year = {2026},
}

@article{daryantoConversateSupportingReflective2025,
    title = {Conversate: {Supporting} {Reflective} {Learning} in {Interview} {Practice} {Through} {Interactive} {Simulation} and {Dialogic} {Feedback}},
    volume = {9},
    shorttitle = {Conversate},
    url = {https://dl.acm.org/doi/10.1145/3701188},
    doi = {10.1145/3701188},
    number = {1},
    urldate = {2026-08-29},
    journal = {Proceedings of the ACM on Human-Computer Interaction},
    author = {Daryanto, Taufiq and Ding, Xiaohan and Wilhelm, Lance T. and Stil, Sophia and Knutsen, Kirk McInnis and Rho, Eugenia H.},
    month = jan,
    year = {2025},
    pages = {GROUP9:1--GROUP9:32},
}

@article{tangGenerativeAIDialogic2026,
    title = {Generative {AI} as a {Dialogic} {Partner}: {Enhancing} {Multiple} {Perspectives}, {Reasoning}, and {Argumentation} in {Science} {Education} with {Customized} {Chatbots}},
    volume = {35},
    issn = {1573-1839},
    shorttitle = {Generative {AI} as a {Dialogic} {Partner}},
    url = {https://doi.org/10.1007/s10956-025-10240-1},
    doi = {10.1007/s10956-025-10240-1},
    language = {en},
    number = {1},
    urldate = {2026-08-29},
    journal = {Journal of Science Education and Technology},
    author = {Tang, Kok-Sing and Putra, Gde Buana Sandila},
    month = feb,
    year = {2026},
    pages = {128--140},
}

@article{chiLearningHumanTutoring2001,
    title = {Learning from human tutoring},
    volume = {25},
    copyright = {© 2001 Cognitive Science Society, Inc.},
    issn = {1551-6709},
    url = {https://onlinelibrary.wiley.com/doi/abs/10.1207/s15516709cog2504_1},
    doi = {10.1207/s15516709cog2504_1},
    language = {en},
    number = {4},
    urldate = {2026-08-29},
    journal = {Cognitive Science},
    author = {Chi, Michelene T.H. and Siler, Stephanie A. and Jeong, Heisawn and Yamauchi, Takashi and Hausmann, Robert G.},
    year = {2001},
    pages = {471--533},
}

@inproceedings{chenIntegratingAITeaching2025,
    address = {New York, NY, USA},
    series = {{DEIT} '25},
    title = {Integrating {AI} {Teaching} {Assistants} and {Knowledge} {Graphs} in {College} {English} {Listening} and {Speaking} {Education}: {A} {Study} of {Evaluation} and {Practice}},
    isbn = {979-8-4007-1324-8},
    shorttitle = {Integrating {AI} {Teaching} {Assistants} and {Knowledge} {Graphs} in {College} {English} {Listening} and {Speaking} {Education}},
    url = {https://dl.acm.org/doi/10.1145/3732299.3732316},
    doi = {10.1145/3732299.3732316},
    urldate = {2026-08-29},
    booktitle = {Proceedings of the 2025 {International} {Conference} on {Digital} {Education} and {Information} {Technology}},
    publisher = {Association for Computing Machinery},
    author = {Chen, Tianhui and Gong, Lanlan and Wang, Ning and Ling, Xinghong},
    month = jun,
    year = {2025},
    pages = {83--90},
}

@inproceedings{zhouResearchApplicationDigital2024,
    address = {New York, NY, USA},
    series = {{ICDEL} '24},
    title = {Research on the {Application} of {Digital} {Humans} in {English} {Oral} {Teaching} {Based} on {AI} {Models}},
    isbn = {979-8-4007-1680-5},
    url = {https://dl.acm.org/doi/10.1145/3675812.3675820},
    doi = {10.1145/3675812.3675820},
    urldate = {2026-08-29},
    booktitle = {Proceedings of the 2024 9th {International} {Conference} on {Distance} {Education} and {Learning}},
    publisher = {Association for Computing Machinery},
    author = {Zhou, YuJie and Xu, Ke and Yin, BoQiao and Liu, NingNing},
    month = aug,
    year = {2024},
    pages = {49--56},
}

@inproceedings{sharmaComuniqaExploringLarge2024,
    address = {New York, NY, USA},
    series = {{COMPASS} '24},
    title = {Comuniqa: {Exploring} {Large} {Language} {Models} {For} {Improving} {English} {Speaking} {Skills}},
    isbn = {979-8-4007-1048-3},
    shorttitle = {Comuniqa},
    url = {https://dl.acm.org/doi/10.1145/3674829.3675082},
    doi = {10.1145/3674829.3675082},
    urldate = {2026-08-29},
    booktitle = {Proceedings of the 7th {ACM} {SIGCAS}/{SIGCHI} {Conference} on {Computing} and {Sustainable} {Societies}},
    publisher = {Association for Computing Machinery},
    author = {Sharma, Shikhar and Mhasakar, Manas and Mehra, Apurv and Venaik, Utkarsh and Singhal, Ujjwal and Kumar, Dhruv and Mittal, Kashish},
    month = aug,
    year = {2024},
    pages = {256--267},
}

@inproceedings{parkAITwinEnhancing2026,
    address = {New York, NY, USA},
    series = {{CHI} '26},
    title = {{AI} {Twin}: {Enhancing} {ESL} {Speaking} {Practice} through {AI} {Self}-{Clones} of a {Better} {Me}},
    isbn = {979-8-4007-2278-3},
    shorttitle = {{AI} {Twin}},
    url = {https://dl.acm.org/doi/10.1145/3772318.3790266},
    doi = {10.1145/3772318.3790266},
    urldate = {2026-08-29},
    booktitle = {Proceedings of the 2026 {CHI} {Conference} on {Human} {Factors} in {Computing} {Systems}},
    publisher = {Association for Computing Machinery},
    author = {Park, Minju and Lee, Seunghyun and Ma, Juhwan and Yoon, Dongwook},
    month = apr,
    year = {2026},
    pages = {1--21},
}

@inproceedings{dejesusCognitiveLoadSupporting2025,
    address = {New York, NY, USA},
    series = {{RESPECT} 2025},
    title = {Cognitive {Load} and {Supporting} {Multilingual} {Learners} in {Computer} {Science} {K}-12 {Classrooms}},
    isbn = {979-8-4007-1355-2},
    url = {https://dl.acm.org/doi/10.1145/3704637.3734736},
    doi = {10.1145/3704637.3734736},
    urldate = {2026-08-29},
    booktitle = {Proceedings of the 2025 {Conference} on {Research} on {Equitable} and {Sustained} {Participation} in {Engineering}, {Computing}, and {Technology}},
    publisher = {Association for Computing Machinery},
    author = {De Jesús, Sofía and Barnes, Timothy and Stehlik, Mark and Kosbie, David and Bozzo Cawley, Erin},
    month = jul,
    year = {2025},
    pages = {13--17},
}

@article{dejongCognitiveLoadTheory2010,
    title = {Cognitive load theory, educational research, and instructional design: some food for thought},
    volume = {38},
    issn = {1573-1952},
    shorttitle = {Cognitive load theory, educational research, and instructional design},
    url = {https://doi.org/10.1007/s11251-009-9110-0},
    doi = {10.1007/s11251-009-9110-0},
    language = {en},
    number = {2},
    urldate = {2026-08-30},
    journal = {Instructional Science},
    author = {de Jong, Ton},
    month = mar,
    year = {2010},
    pages = {105--134},
}

@article{ackermanPromiseWritingLearn1993,
    title = {The {Promise} of {Writing} to {Learn}},
    volume = {10},
    issn = {0741-0883},
    url = {https://doi.org/10.1177/0741088393010003002},
    doi = {10.1177/0741088393010003002},
    language = {EN},
    number = {3},
    urldate = {2026-08-30},
    journal = {Written Communication},
    publisher = {SAGE Publications Inc},
    author = {ACKERMAN, JOHN M.},
    month = jul,
    year = {1993},
    pages = {334--370},
}

@article{kleinReopeningInquiryCognitive1999,
    title = {Reopening {Inquiry} into {Cognitive} {Processes} in {Writing}-{To}-{Learn}},
    volume = {11},
    issn = {1573-336X},
    url = {https://doi.org/10.1023/A:1021913217147},
    doi = {10.1023/A:1021913217147},
    language = {en},
    number = {3},
    urldate = {2026-08-30},
    journal = {Educational Psychology Review},
    author = {Klein, Perry D.},
    month = sep,
    year = {1999},
    pages = {203--270},
}

@inproceedings{marinaDetectionRethinkingEducation2025,
    address = {Cham},
    title = {Beyond {Detection}: {Rethinking} {Education} in the {Age} of {AI}-{Writing}},
    isbn = {978-3-031-99267-4},
    shorttitle = {Beyond {Detection}},
    doi = {10.1007/978-3-031-99267-4_1},
    language = {en},
    booktitle = {Artificial {Intelligence} in {Education}. {Posters} and {Late} {Breaking} {Results}, {Workshops} and {Tutorials}, {Industry} and {Innovation} {Tracks}, {Practitioners}, {Doctoral} {Consortium}, {Blue} {Sky}, and {WideAIED}},
    publisher = {Springer Nature Switzerland},
    author = {Marina, Maria and Panchenko, Alexander and Konovalov, Vasily},
    editor = {Cristea, Alexandra I. and Walker, Erin and Lu, Yu and Santos, Olga C. and Isotani, Seiji},
    year = {2025},
    pages = {3--10},
}

@article{wattsComparingStudentGenerative2023,
    title = {Comparing {Student} and {Generative} {Artificial} {Intelligence} {Chatbot} {Responses} to {Organic} {Chemistry} {Writing}-to-{Learn} {Assignments}},
    volume = {100},
    issn = {0021-9584},
    url = {https://doi.org/10.1021/acs.jchemed.3c00664},
    doi = {10.1021/acs.jchemed.3c00664},
    number = {10},
    urldate = {2026-08-30},
    journal = {Journal of Chemical Education},
    author = {Watts, Field M. and Dood, Amber J. and Shultz, Ginger V. and Rodriguez, Jon-Marc G.},
    month = sep,
    year = {2023},
    pages = {3806--3817},
}

@article{kimEmpoweringDiverseLearners2024,
    title = {Empowering {Diverse} {Learners}: {Integrating} {Writing}-to-{Learn} {Strategies} in a {Middle} {School} {Science} {Classroom} in the {U}.{S}.},
    volume = {14},
    copyright = {http://creativecommons.org/licenses/by/3.0/},
    issn = {2227-7102},
    shorttitle = {Empowering {Diverse} {Learners}},
    url = {https://www.mdpi.com/2227-7102/14/9/1031},
    doi = {10.3390/educsci14091031},
    language = {en},
    number = {9},
    urldate = {2026-08-30},
    journal = {Education Sciences},
    publisher = {Multidisciplinary Digital Publishing Institute},
    author = {Kim, So Lim and Kim, Deoksoon},
    month = sep,
    year = {2024},
    pages = {1031},
}

@article{graesserCollaborativeDialoguePatterns1995,
    title = {Collaborative dialogue patterns in naturalistic one-to-one tutoring},
    volume = {9},
    copyright = {Copyright © 1995 John Wiley \& Sons, Ltd},
    issn = {1099-0720},
    url = {https://onlinelibrary.wiley.com/doi/abs/10.1002/acp.2350090604},
    doi = {10.1002/acp.2350090604},
    language = {en},
    number = {6},
    urldate = {2026-08-30},
    journal = {Applied Cognitive Psychology},
    author = {Graesser, Arthur C. and Person, Natalie K. and Magliano, Joseph P.},
    year = {1995},
    pages = {495--522},
}

@article{paladinesSystematicLiteratureReview2020,
    title = {A {Systematic} {Literature} {Review} of {Intelligent} {Tutoring} {Systems} {With} {Dialogue} in {Natural} {Language}},
    volume = {8},
    issn = {2169-3536},
    url = {https://ieeexplore.ieee.org/abstract/document/9186073},
    doi = {10.1109/ACCESS.2020.3021383},
    urldate = {2026-08-30},
    journal = {IEEE Access},
    author = {Paladines, José and Ramirez, Jaime},
    year = {2020},
    pages = {164246--164267},
}

@misc{bealeDialogicPedagogyLarge2025,
    title = {Dialogic {Pedagogy} for {Large} {Language} {Models}: {Aligning} {Conversational} {AI} with {Proven} {Theories} of {Learning}},
    shorttitle = {Dialogic {Pedagogy} for {Large} {Language} {Models}},
    url = {http://arxiv.org/abs/2506.19484},
    doi = {10.48550/arXiv.2506.19484},
    urldate = {2026-08-30},
    publisher = {arXiv},
    author = {Beale, Russell},
    month = jun,
    year = {2025},
    note = {arXiv:2506.19484 [cs.CL]},
}

@inproceedings{liuSingaKidsMultilingualMultimodal2025,
    address = {Vienna, Austria},
    title = {{SingaKids}: {A} {Multilingual} {Multimodal} {Dialogic} {Tutor} for {Language} {Learning}},
    isbn = {979-8-89176-288-6},
    shorttitle = {{SingaKids}},
    url = {https://aclanthology.org/2025.acl-industry.86/},
    doi = {10.18653/v1/2025.acl-industry.86},
    urldate = {2026-05-11},
    booktitle = {Proceedings of the 63rd {Annual} {Meeting} of the {Association} for {Computational} {Linguistics} ({Volume} 6: {Industry} {Track})},
    publisher = {Association for Computational Linguistics},
    author = {Liu, Zhengyuan and Lin, Geyu and Tan, Hui Li and Zhang, Huayun and Lu, Yanfeng and Gao, Xiaoxue and Yin, Stella Xin and He, Sun and Goh, Hock Huan and Wong, Lung Hsiang and Chen, Nancy F.},
    editor = {Rehm, Georg and Li, Yunyao},
    month = jul,
    year = {2025},
    pages = {1244--1253},
}

@inproceedings{tianSystematicLiteratureReview2025,
    address = {New York, NY, USA},
    series = {{AISNS} '24},
    title = {A {Systematic} {Literature} {Review} on the {Feasibility} of {Dialogue} {Learning} {Based} on {Generative} {AI}},
    isbn = {979-8-4007-1123-7},
    url = {https://dl.acm.org/doi/10.1145/3714334.3714362},
    doi = {10.1145/3714334.3714362},
    urldate = {2026-08-30},
    booktitle = {Proceedings of the 2024 2nd {International} {Conference} on {Artificial} {Intelligence}, {Systems} and {Network} {Security}},
    publisher = {Association for Computing Machinery},
    author = {Tian, Yusen and Tu, Binghao},
    month = mar,
    year = {2025},
    pages = {160--164},
}

@inproceedings{liuAmIWriteExploringScalable2026,
    address = {New York, NY, USA},
    series = {{CHI} '26},
    title = {{AmIWrite}: {Exploring} {Scalable} {One}-on-{One} {Handwriting}-{Based} {Tutoring} for {Mathematical} {Problem}-{Solving} with an {LLM}-{Powered} {AI} {Tutor}},
    isbn = {979-8-4007-2278-3},
    shorttitle = {{AmIWrite}},
    url = {https://dl.acm.org/doi/10.1145/3772318.3790935},
    doi = {10.1145/3772318.3790935},
    urldate = {2026-08-30},
    booktitle = {Proceedings of the 2026 {CHI} {Conference} on {Human} {Factors} in {Computing} {Systems}},
    publisher = {Association for Computing Machinery},
    author = {Liu, Ziyi and Chen, Yuzhao and Ji, Haoyu and Duan, Runlin and Zhu, Zhengzhe and Hu, Xiyun and Peppler, Kylie and Ramani, Karthik},
    month = apr,
    year = {2026},
    pages = {1--28},
}

@incollection{renshawDialogicLearningTeaching2004,
    address = {Dordrecht},
    title = {Dialogic {Learning} {Teaching} and {Instruction}},
    isbn = {978-1-4020-1931-9},
    url = {https://doi.org/10.1007/1-4020-1931-9_1},
    doi = {10.1007/1-4020-1931-9_1},
    language = {en},
    urldate = {2026-09-06},
    booktitle = {Dialogic {Learning}: {Shifting} {Perspectives} to {Learning}, {Instruction}, and {Teaching}},
    publisher = {Springer Netherlands},
    author = {Renshaw, Peter D.},
    editor = {van der Linden, Jos and Renshaw, Peter},
    year = {2004},
    pages = {1--15},
}

@incollection{clarkGroundingCommunication1991,
    address = {Washington, DC, US},
    title = {Grounding in communication},
    isbn = {978-1-55798-121-9},
    doi = {10.1037/10096-006},
    booktitle = {Perspectives on socially shared cognition},
    publisher = {American Psychological Association},
    author = {Clark, Herbert H. and Brennan, Susan E.},
    year = {1991},
    pages = {127--149},
}

@article{macaroSystematicReviewEnglish2018,
    title = {A systematic review of {English} medium instruction in higher education},
    volume = {51},
    issn = {0261-4448, 1475-3049},
    url = {https://www.cambridge.org/core/journals/language-teaching/article/systematic-review-of-english-medium-instruction-in-higher-education/E802DA0854E0726F3DE213548B7B7EC7},
    doi = {10.1017/S0261444817000350},
    language = {en},
    number = {1},
    urldate = {2026-09-11},
    journal = {Language Teaching},
    author = {Macaro, Ernesto and Curle, Samantha and Pun, Jack and An, Jiangshan and Dearden, Julie},
    month = jan,
    year = {2018},
    pages = {36--76},
}

@article{fenton-smithUnpackingEnglishMedium2026,
    title = {Unpacking the {English} {Medium} {Instruction} ({EMI}) boom: exploring student enrolment and driving forces in {China} and {Japan}},
    volume = {0},
    issn = {0267-1522},
    shorttitle = {Unpacking the {English} {Medium} {Instruction} ({EMI}) boom},
    url = {https://doi.org/10.1080/02671522.2026.2633628},
    doi = {10.1080/02671522.2026.2633628},
    number = {0},
    urldate = {2026-09-11},
    journal = {Research Papers in Education},
    publisher = {Routledge},
    author = {Fenton-Smith, Ben and Zhou, Jingwen and Galloway, Nicola},
    month = feb,
    year = {2026},
    pages = {1--25},
}

@article{alhazmiImpactBilingualVs2025a,
  title = {The {{Impact}} of {{Bilingual}} vs {{English-Only Instruction}} on the {{Performance}} of {{Undergraduate Saudi Medical Science Students}}},
  author = {Alhazmi, Aiman and Alzahrani, Mohammed},
  date = {2025-06},
  journaltitle = {Advances in Medical Education and Practice},
  shortjournal = {AMEP},
  volume = {Volume 16},
  pages = {1063--1075},
  issn = {1179-7258},
  doi = {10.2147/AMEP.S520706},
  url = {https://www.dovepress.com/the-impact-of-bilingual-vs-english-only-instruction-on-the-performance-peer-reviewed-fulltext-article-AMEP},
  urldate = {2026-02-20},
  langid = {english}
}

@article{azamReturnsEnglishLanguageSkills,
  title = {The {{Returns}} to {{English-Language Skills}} in {{India}}},
  author = {Azam, Mehtabul and Chin, Aimee and Prakash, Nishith},
  langid = {english}
}

@article{bhattacharyaMediatingInequalitiesExploring2013,
  title = {Mediating Inequalities: Exploring {{English-medium}} Instruction in a Suburban {{Indian}} Village School},
  shorttitle = {Mediating Inequalities},
  author = {Bhattacharya, Usree},
  date = {2013-02-01},
  journaltitle = {Current Issues in Language Planning},
  volume = {14},
  number = {1},
  pages = {164--184},
  publisher = {Routledge},
  issn = {1466-4208},
  doi = {10.1080/14664208.2013.791236},
  url = {https://doi.org/10.1080/14664208.2013.791236},
  urldate = {2026-04-30}
}

@article{brisk2012challenges,
  title = {Challenges and Supports for {{English}} Language Learners in Bilingual Programs},
  author = {Brisk, Mar{\'i}a Estela and Proctor, C Patrick},
  date = {2012},
  journaltitle = {Commissioned Papers on Language and Literacy Issues in the Common Core State Standards and Next Generation Science Standards},
  volume = {94},
  pages = {115},
  publisher = {Stanford University Stanford, CA}
}

@article{chirikovGenerativeAIUse2026,
  title = {Generative {{AI}} Use and Misuse Call for Assessment Reform in Higher Education},
  author = {Chirikov, Igor and Smirnov, Ivan and Kizilcec, René F.},
  date = {2026-05-21},
  journaltitle = {Science},
  shortjournal = {Science},
  volume = {392},
  number = {6800},
  pages = {818--820},
  issn = {0036-8075, 1095-9203},
  doi = {10.1126/science.aec5115},
  url = {https://www.science.org/doi/10.1126/science.aec5115},
  urldate = {2026-07-07},
  langid = {english}
}

@article{civanEffectMediumInstruction2016,
  title = {The {{Effect}} of the {{Medium}} of {{Instruction Language}} on the {{Academic Success}} of {{University Students}}},
  author = {Civan, Abdulkadir and Coskun, Ali},
  date = {2016-12},
  journaltitle = {Educational Sciences: Theory and Practice},
  volume = {16},
  number = {6},
  pages = {1981--2004},
  publisher = {Educational Consultancy, Ltd (EDAM)},
  issn = {1303-0485},
  url = {https://eric.ed.gov/?id=EJ1130741},
  urldate = {2026-02-20},
  langid = {english}
}

@article{karabayEvolutionEnglishMedium2024,
  title = {The {{Evolution}} of {{English Medium Instruction Research}} in {{Higher Education}}: {{A Bibliometric Study}}},
  shorttitle = {The {{Evolution}} of {{English Medium Instruction Research}} in {{Higher Education}}},
  author = {Karabay, Akmaral and Durrani, Naureen},
  date = {2024-09-05},
  journaltitle = {Education Sciences},
  shortjournal = {Education Sciences},
  volume = {14},
  number = {9},
  pages = {982},
  issn = {2227-7102},
  doi = {10.3390/educsci14090982},
  url = {https://www.mdpi.com/2227-7102/14/9/982},
  urldate = {2026-07-26},
  langid = {english}
}

@article{khanEnglishLinguisticCapital2025,
  title = {English {{Linguistic Capital}} and {{Career Trajectory}}: {{Impact}} of {{Medium}} of {{Instruction}} on the {{Professional Life}} of {{Indians}}},
  shorttitle = {English {{Linguistic Capital}} and {{Career Trajectory}}},
  author = {Khan, Sadaf and Jayaraj, Thapasya},
  date = {2025-01-10},
  journaltitle = {Journal of Language, Identity \& Education},
  volume = {0},
  number = {0},
  pages = {1--14},
  publisher = {Routledge},
  issn = {1534-8458},
  doi = {10.1080/15348458.2024.2444280},
  url = {https://doi.org/10.1080/15348458.2024.2444280},
  urldate = {2026-02-20}
}

@article{maInvestigationImpactCognitive2025,
  title = {Investigation of the Impact of Cognitive Load on {{EFL}} Learners’ Satisfaction with {{MOOCs}}: The Mediating Role of Expectation Confirmation and Perceived Usefulness},
  shorttitle = {Investigation of the Impact of Cognitive Load on {{EFL}} Learners’ Satisfaction with {{MOOCs}}},
  author = {Ma, Lingling},
  date = {2025-04-22},
  journaltitle = {BMC Psychology},
  shortjournal = {BMC Psychol},
  volume = {13},
  number = {1},
  pages = {416},
  issn = {2050-7283},
  doi = {10.1186/s40359-025-02735-8},
  url = {https://bmcpsychology.biomedcentral.com/articles/10.1186/s40359-025-02735-8},
  urldate = {2026-07-26},
  langid = {english}
}

@article{meleTalkingVsTyping2026,
  title = {Talking vs Typing: How Voice- vs Text-Based Educational Robots Shape Student Engagement},
  shorttitle = {Talking vs Typing},
  author = {Mele, Cristina and Russo-Spena, Tiziana and Ranieri, Angelo and Di Bernardo, Irene and Hollebeek, Linda D.},
  date = {2026-02-06},
  journaltitle = {Journal of Service Management},
  volume = {37},
  number = {1},
  pages = {141--164},
  issn = {1757-5818, 1757-5826},
  doi = {10.1108/JOSM-02-2024-0082},
  url = {https://www.emerald.com/josm/article/37/1/141/1278806/Talking-vs-typing-how-voice-vs-text-based},
  urldate = {2026-03-08},
  langid = {english}
}

@article{rocabadoYouCanStand2025,
  title = {You {{Can Stand Under My Umbrella}}: {{Cognitive Load}} in {{Second-Language Reading}}},
  shorttitle = {You {{Can Stand Under My Umbrella}}},
  author = {Rocabado, Francisco and Schmitz, Gianna and Duñabeitia, Jon Andoni},
  date = {2025-08-03},
  journaltitle = {Behavioral Sciences},
  shortjournal = {Behavioral Sciences},
  volume = {15},
  number = {8},
  pages = {1051},
  issn = {2076-328X},
  doi = {10.3390/bs15081051},
  url = {https://www.mdpi.com/2076-328X/15/8/1051},
  urldate = {2026-07-26},
  langid = {english}
}

@article{rousselLearningSubjectContent2017a,
  title = {Learning Subject Content through a Foreign Language Should Not Ignore Human Cognitive Architecture: {{A}} Cognitive Load Theory Approach},
  shorttitle = {Learning Subject Content through a Foreign Language Should Not Ignore Human Cognitive Architecture},
  author = {Roussel, Stéphanie and Joulia, Danielle and Tricot, André and Sweller, John},
  date = {2017-12},
  journaltitle = {Learning and Instruction},
  shortjournal = {Learning and Instruction},
  volume = {52},
  pages = {69--79},
  issn = {09594752},
  doi = {10.1016/j.learninstruc.2017.04.007},
  url = {https://linkinghub.elsevier.com/retrieve/pii/S0959475216302584},
  urldate = {2026-02-20},
  langid = {english}
}

@article{rousselAdvantagesListeningAcademic2022,
  title = {The Advantages of Listening to Academic Content in a Second Language May Be Outweighed by Disadvantages: {{A}} Cognitive Load Theory Approach},
  shorttitle = {The Advantages of Listening to Academic Content in a Second Language May Be Outweighed by Disadvantages},
  author = {Roussel, Stéphanie and Tricot, André and Sweller, John},
  date = {2022},
  journaltitle = {British Journal of Educational Psychology},
  volume = {92},
  number = {2},
  pages = {627--644},
  issn = {2044-8279},
  doi = {10.1111/bjep.12468},
  url = {https://onlinelibrary.wiley.com/doi/abs/10.1111/bjep.12468},
  urldate = {2026-02-20},
  langid = {english}
}

@incollection{swellerCognitiveLoadTheory2011,
  title = {Cognitive {{Load Theory}}},
  booktitle = {Psychology of {{Learning}} and {{Motivation}}},
  author = {Sweller, John},
  date = {2011},
  volume = {55},
  pages = {37--76},
  publisher = {Elsevier},
  doi = {10.1016/B978-0-12-387691-1.00002-8},
  url = {https://linkinghub.elsevier.com/retrieve/pii/B9780123876911000028},
  urldate = {2026-02-24},
  isbn = {978-0-12-387691-1},
  langid = {english}
}

@online{thakurEnglishMediumBoom2025,
  title = {English Medium Boom Risks Education and Identity | {{Policy Circle}}},
  author = {Thakur, Debdulal},
  date = {2025-08-01T07:00:31+05:30},
  url = {https://www.policycircle.org/opinion/english-medium-education-nep/},
  urldate = {2026-04-30},
  langid = {british}
}

@article{treffers-dallerHowReadyAre2022,
  title = {How {{Ready Are Indian Primary School Children}} for {{English Medium Instruction}}? {{An Analysis}} of the {{Relationship}} between the {{Reading Skills}} of {{Low-SES Children}}, {{Their Oral Vocabulary}} and {{English Input}} in the {{Classroom}} in {{Government Schools}} in {{India}}},
  shorttitle = {How {{Ready Are Indian Primary School Children}} for {{English Medium Instruction}}?},
  author = {Treffers-Daller, Jeanine and Mukhopadhyay, Lina and Balasubramanian, Anusha and Tamboli, Vasim and Tsimpli, Ianthi},
  date = {2022-08-01},
  journaltitle = {Applied Linguistics},
  shortjournal = {Appl Linguist},
  volume = {43},
  number = {4},
  pages = {746--775},
  issn = {0142-6001},
  doi = {10.1093/applin/amac003},
  url = {https://doi.org/10.1093/applin/amac003},
  urldate = {2026-04-30}
}

@article{koedinger2001cognitive,
  title={Cognitive tutors},
  author={Koedinger, Kenneth R and Corbett, Albert},
  journal={Smart machines in education},
  pages={145--167},
  year={2001},
  publisher={Cambridge, MA: MIT Press}
}

\appendix


\section{Appendix}
\subsection{Learning Activity Interfaces}\label{app:interface-images}
\begin{figure*}[t]
    \centering
    \begin{minipage}[t]{0.31\textwidth}
        \centering
        \includegraphics[width=\linewidth]{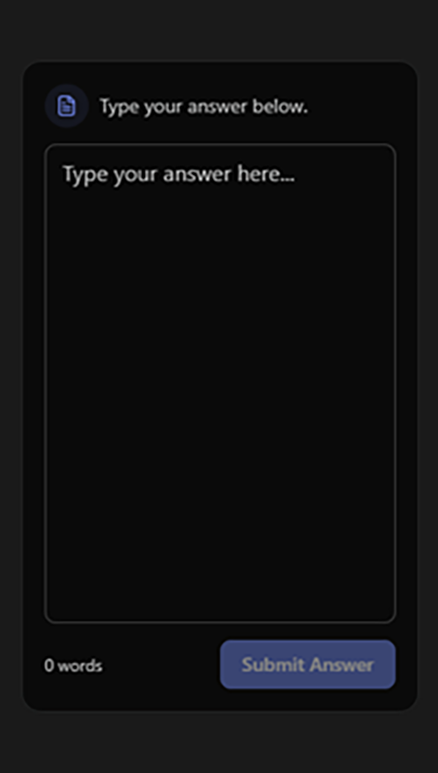}
        \par\smallskip
        \textbf{(a) Written response}
    \end{minipage}
    \hfill
    \begin{minipage}[t]{0.31\textwidth}
        \centering
        \includegraphics[width=\linewidth]{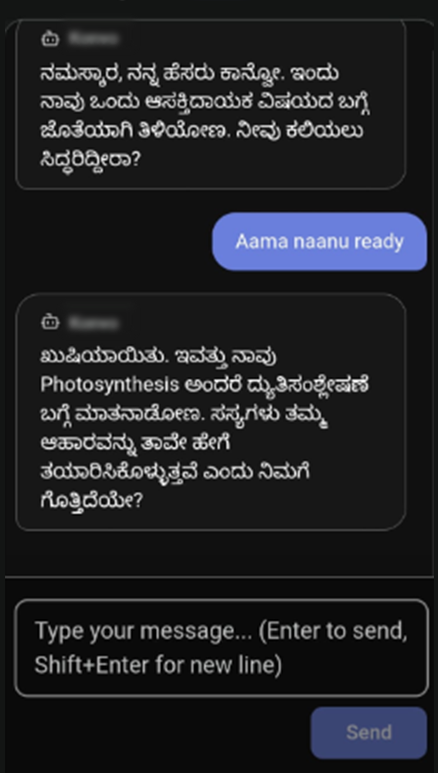}
        \par\smallskip
        \textbf{(b) Text dialogue}
    \end{minipage}
    \hfill
    \begin{minipage}[t]{0.31\textwidth}
        \centering
        \includegraphics[width=\linewidth]{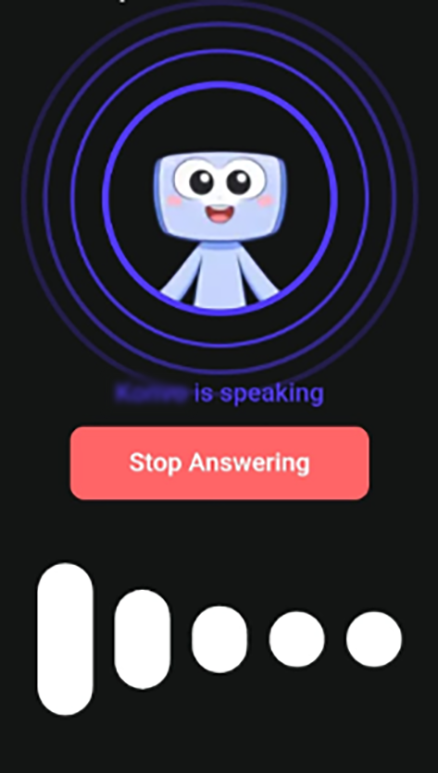}
        \par\smallskip
        \textbf{(c) Voice dialogue}
    \end{minipage}
    \caption{Interfaces for the three activity formats: (a) written response, (b) text dialogue with the AI tutor, and (c) voice dialogue with the AI tutor.}
    \Description{Three mobile interface screenshots. The written-response interface contains a text-entry box and submit button. The text-dialogue interface shows a bilingual text conversation with the AI tutor and a message-entry box. The voice-dialogue interface shows an animated tutor, a button for stopping the tutor's response, and an audio waveform.}
    \label{fig:activity-interfaces}
\end{figure*}

Refer to Figure~\ref{fig:activity-interfaces}

\subsection{Learning Module}\label{apx:learning-module}
We developed the learning module in collaboration with the implementation partner's pedagogy team. The module introduced data-driven reasoning---how digital systems use data to produce conclusions, suggestions, and other outputs---through four instructional videos:

\begin{itemize}
    \item \textit{Introduction to Data-Driven Reasoning} introduced how digital systems use patterns in data to produce outputs and how those outputs reflect the systems' data, design choices, and goals.

    \item \textit{Predictive Reasoning} explained how systems use past data to estimate future or unknown outcomes and why such predictions express likelihood rather than certainty.

    \item \textit{Discovery Reasoning} described how systems identify patterns or groupings in data and why the resulting structures depend on the data and features selected.

    \item \textit{Generative Reasoning} explained how systems create new content from learned patterns and why plausible or realistic outputs still require human verification.
\end{itemize}

Each video was five to eight minutes long. Consistent with the instructional practices used in participants' existing AI literacy program, the videos presented visual content and technical terminology in English, while the narration combined Kannada and English. Each video was followed by a corresponding learning activity. The Study Methodology section describes the activities, experimental conditions, and study platform; we do not repeat those details here.

\subsection{Survey Instruments} \label{apx:surveys}
This appendix reproduces the pre-survey and the post-survey. Both surveys were prepared in two forms, A and B, with the questions randomized in different orders. We reproduce Form A here.

\subsubsection{Pre-Survey}

\textbf{Section 1: Demographic Information.} Name, phone, degree, and year.

\textbf{Section 2: Awareness.} Students rated their awareness of the following topics on a five-point scale from ``Not at all aware'' to ``Extremely aware'': (i) Predictive Reasoning, (ii) Discovery Reasoning, (iii) Generative Reasoning.

\textbf{Section 3: AI Knowledge.} Students answered the following 17 questions without external help.

\begin{enumerate}
  \item Which of the following scenarios is an example of a regression task?
  \begin{enumerate}
    \item Predicting that a student will score 68 marks on an exam
    \item Labeling emails as ``spam'' or ``not spam''
    \item Grouping movies into categories based on viewer preferences
    \item Translating a sentence from English to Tamil
  \end{enumerate}
  \item Which of the following scenarios is an example of a classification task?
  \begin{enumerate}
    \item Predicting the price of a house based on its size and location
    \item Labeling emails as ``spam'' or ``not spam''
    \item Grouping customers into clusters based on purchasing behavior
    \item Breaking a sentence into individual words
  \end{enumerate}
  \item An AI system analyzes past loan records to predict whether new customers will repay their loans. What allows it to make these predictions?
  \begin{enumerate}
    \item It finds patterns in past examples and uses them for new cases
    \item It uses rules to decide
    \item It guesses based on chance
    \item It relies on humans to make the prediction
  \end{enumerate}
  \item Do AI systems always make accurate predictions? Why or why not? (open-ended)
  \item Google Maps shows that it will take about 48 minutes to travel from JP Nagar to Cubbon Park. What factors does Google Maps consider, and how does it use them to predict the travel time? (open-ended)
  \item What is the difference between supervised and unsupervised learning regarding labeled data? (open-ended)
  \item Which of the following is an example of an Unsupervised Learning task?
  \begin{enumerate}
    \item Classification
    \item Regression
    \item Clustering
    \item Generation
  \end{enumerate}
  \item Which of the following scenarios is an example of clustering?
  \begin{enumerate}
    \item Predicting tomorrow's temperature
    \item Grouping songs based on similar sound patterns
    \item Classifying medical images as healthy or diseased
    \item Translating text from Telugu to English
  \end{enumerate}
  \item A music streaming app groups users based on the type of music they listen to (for example, hip-hop or classical). The app then recommends songs that are popular in each group. How is clustering helping the recommendation system?
  \begin{enumerate}
    \item By grouping users with similar music preferences
    \item By predicting the popular song
    \item By deleting old songs
    \item By choosing songs randomly
  \end{enumerate}
  \item List a few mobile apps you use that recommend videos, products, or personalized ads to you. Explain how the app decides what to recommend. (open-ended)
  \item When you ask an AI chatbot like ChatGPT a question, it generates a response. How is the response generated? (open-ended)
  \item Before generating text, an AI system breaks a sentence into smaller units like words or parts of words. What is this process called?
  \begin{enumerate}
    \item Clustering
    \item Tokenization
    \item Classification
    \item Bias
  \end{enumerate}
  \item Which of the following is an example of hallucination in generative AI?
  \begin{enumerate}
    \item An AI chatbot makes up false information and presents it as a real fact
    \item An AI tool summarizes a document
    \item An AI tool translates a sentence
    \item An AI system predicts exam scores
  \end{enumerate}
  \item An AI system generates a realistic video of a person saying something they never said. What concept best explains this?
  \begin{enumerate}
    \item Classification
    \item Regression
    \item Deep Fakes
    \item Tokenization
  \end{enumerate}
  \item AI chatbots can generate answers to many questions. Does this mean they know everything? Explain why or why not. (open-ended)
\end{enumerate}

\textbf{Section 4: AI Interest and Self-Efficacy.} Students responded to the following items on a five-point Likert scale from ``Strongly disagree'' to ``Strongly agree''.

Q16. Interest in AI:
\begin{enumerate}
  \item I generally have fun when I am learning about AI.
  \item I like reading about AI.
  \item I like to work on problems in the field of AI.
  \item I enjoy acquiring new knowledge about AI.
  \item I am interested in learning new things about AI.
\end{enumerate}

Q17. Confidence with AI (self-efficacy):
\begin{enumerate}
  \item I have a good understanding of the basic principles of AI.
  \item I know about various uses of AI.
  \item I understand what ``intelligence'' means in the context of AI.
  \item I am aware of the strengths and weaknesses of AI.
  \item I am familiar with how AI works.
  \item I understand to what extent humans can influence AI.
  \item I can analyze, illustrate, and critically interpret data.
  \item I am familiar with the ethical issues surrounding AI.
\end{enumerate}

\subsubsection{Post-Survey}
The post-survey repeated Sections 1 through 4 of the pre-survey without change, except that the Section 3 instruction asked students to answer based on their learnings from the course. The post-survey added the following two sections.

\textbf{Section 5: Activities Preference.}

Q18. Compared to your usual learning activities (for example, writing answers as homework), how do you feel about the learning activity in this course?
\begin{enumerate}
  \item I strongly prefer the activity in this course
  \item I prefer the activity in this course
  \item It feels about the same
  \item I prefer my usual assignments
  \item I strongly prefer my usual assignments
\end{enumerate}

Q19. Why did you choose the above option? (open-ended)

\textbf{Section 6: Research Interview.}

Q20. Would you be interested in being interviewed for our research study? (Yes / No)

\subsection{Conversation Analysis Codebook} \label{apx:codebook}
Table~\ref{tab:codebook} presents the final codebook used for the conversation analysis. The codes are grouped by who they apply to. Four codes apply to student turns, four apply to agent turns, and one applies to the conversation as a whole. The table lists each group, its codes, a short description, and the number of units labeled with each code. We report two sets of counts. The gold column gives the counts from the 100 conversations labeled by the second author. The corpus column gives the counts from the full annotated corpus of 1,604 conversations, after excluding the sample activity. The conversation resolution code is applied at the conversation level rather than the turn level.

\begin{table*}[t]
\centering
\caption{Codebook for the conversation analysis. Gold counts are from the 100 human-labeled conversations (838 turns). Corpus counts are from the 1,604 annotated conversations (11,500 turns, of which 5,124 are student turns) that remain after the sample activity is excluded. The codes are not mutually exclusive, so a turn may carry more than one code within its group. Conversation resolution is counted in conversations rather than turns, as the number in which at least one turn was coded as resolved.}
\label{tab:codebook}
\begin{tabularx}{\textwidth}{l l X r r}
\toprule
Group & Code & Description & Gold & Corpus \\
\midrule
\multirow{4}{*}{Student}
  & Demonstrated Understanding & Student acknowledges understanding or explains the concept & 145 & 1{,}776 \\
  & Emerging Understanding & Student asks for clarification while working toward understanding & 63 & 900 \\
  & Confused & Student does not understand, does not know, or gives a wrong response & 81 & 967 \\
  & Student Friction & Student gives an elliptical or unrelated response & 141 & 1{,}382 \\
\addlinespace
\multirow{4}{*}{Agent}
  & Content Delivery & Agent provides explanations, gives examples, or corrects errors & 156 & 2{,}851 \\
  & Elicitation & Agent scaffolds, prompts, hints, or probes student thinking & 362 & 5{,}135 \\
  & Affective & Agent offers emotional support or praise & 96 & 1{,}655 \\
  & Agent Friction & Agent strays from the goal, asks out-of-context questions, or misunderstands & 50 & 622 \\
\addlinespace
Conversation & Resolved & Whether the conversation reached resolution, coded once for the whole conversation & 33 / 100 & 561 / 1{,}604 \\
\bottomrule
\end{tabularx}
\end{table*}

\subsection{Agreement Between the Model and the Human Labels} \label{apx:irr}
The second author labeled 100 conversations by hand, which served as the gold set for validating the automated annotation. The gold set is balanced across the study design. It contains 50 voice and 50 text chat conversations, 50 in Kannada and 50 in English, and 53 from the first deployment and 47 from the college deployment. All 838 gold-labeled turns were matched to the model output on the conversation and turn identifiers, and the two sources agreed on the speaker for every turn.

Table~\ref{tab:irr} reports the agreement for each code. We report the raw agreement and Cohen's $\kappa$, which corrects for the agreement expected by chance. 

\begin{table*}[t]
\centering
\caption{Agreement between the model annotations and the human gold labels on 100 conversations. The eight turn-level codes are evaluated on turns, student codes on student turns and agent codes on agent turns. Conversation resolution is evaluated on whole conversations. The gold and model columns give the share of units that each source marked with the code.}
\label{tab:irr}
\begin{tabular}{l l r r r r r}
\toprule
Code & Unit & $n$ & Gold & Model & Agreement & $\kappa$ \\
\midrule
Student Demonstrated Understanding & Turn & 380 & 37.9\% & 27.6\% & 84.5\% & 0.65 \\
Student Emerging Understanding & Turn & 380 & 16.6\% & 14.7\% & 80.8\% & 0.27 \\
Student Confused & Turn & 380 & 21.1\% & 22.9\% & 81.3\% & 0.46 \\
Student Friction & Turn & 380 & 36.1\% & 33.7\% & 81.8\% & 0.60 \\
Agent Content Delivery & Turn & 458 & 33.4\% & 37.8\% & 81.7\% & 0.60 \\
Agent Elicitation & Turn & 458 & 78.8\% & 80.1\% & 91.7\% & 0.75 \\
Agent Affective & Turn & 458 & 21.0\% & 23.4\% & 87.6\% & 0.64 \\
Agent Friction & Turn & 458 & 9.4\% & 13.3\% & 91.3\% & 0.57 \\
\addlinespace
Conversation Resolved & Conversation & 100 & 33.0\% & 36.0\% & 93.0\% & 0.85 \\
\bottomrule
\end{tabular}
\end{table*}

\subsection{Interview Protocol} \label{apx:interview-protocol}
The interviews were semi-structured. We used the following questions as a guide and asked follow-up questions as needed.

\begin{enumerate}
  \item What are you studying?
  \item How is teaching done in your class, in English or in Kannada?
  \item How was your experience with the data-driven reasoning course? What did you learn that was new?
  \item How was your experience with the activities? What language did you use to interact with the tool?
  \item What did you like?
  \item What did you not like?
  \item How does it compare to your usual activities in your college courses?
  \item Would you be interested in doing similar activities for your courses in college? Why?
  \item Which language would you be interested in doing the activities in?
  \item Would you prefer voice interaction or text interaction?
  \item Do you use AI for any learning purposes?
\end{enumerate}

\subsection{Interview Codebook} \label{apx:interview-codebook}
Table~\ref{tab:interview-codebook} presents the final codebook used for the interview analysis. The table lists each theme, its codes, and the number of coded instances for each code. The counts are the number of quotations labeled with each code across the 12 transcribed interviews.

\begin{table*}[t]
\centering
\caption{Codebook for the interview analysis. Counts are the number of coded instances for each code across the interviews.}
\label{tab:interview-codebook}
\begin{tabularx}{\textwidth}{l X r}
\toprule
Theme & Code & Count \\
\midrule
\multirow{6}{*}{Patterns of AI Use}
  & AI for assignment production & 16 \\
  & AI easier than textbooks and teachers & 13 \\
  & Attempt first, then AI & 10 \\
  & AI for clarifying doubts and reference & 10 \\
  & Copy-paste without engagement & 9 \\
  & Non-academic AI use & 7 \\
\addlinespace
\multirow{6}{*}{Experience of the AI Tutor}
  & Helpful pedagogical moves & 17 \\
  & Encouraging and non-judgmental & 13 \\
  & Natural conversation quality & 11 \\
  & Fast and responsive & 9 \\
  & Comprehension breakdowns & 6 \\
  & Bot limitations & 3 \\
\addlinespace
\multirow{6}{*}{Language Preferences and Trade-offs}
  & Practising English with the bot & 9 \\
  & English for future opportunity & 7 \\
  & English aligns with curriculum and exams & 5 \\
  & Kannada aids comprehension & 5 \\
  & Need for multilingual support & 2 \\
  & Beyond the Kannada English binary & 2 \\
\addlinespace
\multirow{6}{*}{Modality Preferences and Trade-offs}
  & Interactive activities preferred to written work & 11 \\
  & Text allows review and correction & 8 \\
  & Voice preserves conversational flow & 5 \\
  & Voice burdens memory and attention & 4 \\
  & Text avoids spoken English difficulty & 4 \\
  & Typing barriers in Kannada script & 3 \\
\addlinespace
\multirow{3}{*}{Perceived Effects of AI on Learning}
  & Reduced thinking and over-reliance & 12 \\
  & Perceived prevalence of AI use & 6 \\
  & Academic costs at exam time & 3 \\
\bottomrule
\end{tabularx}
\end{table*}


\end{document}